\pdfoutput=1
\documentclass[12pt,a4]{article}
\usepackage[utf8]{inputenc}
\usepackage[T1]{fontenc}
\usepackage{amsmath}
\usepackage{amsfonts}
\usepackage{amssymb}
\usepackage{graphicx}
\usepackage{fullpage}
\usepackage{hyperref}
\usepackage{tikz}
\usepackage{xparse}
\usepackage[nosort]{cite}

\tikzset{
	dot/.style={circle,fill=black,inner sep=0pt,minimum size=.1cm} }

\usepackage{MnSymbol}
\hypersetup{
	colorlinks=true,
	linkcolor=blue,
	filecolor=magenta,      
	urlcolor=blue,
}

\def\cM{{\mathcal M}}
\def\cN{{\mathcal N}}
\def\cO{{\mathcal O}}

\def\cP{{\mathcal P}}

\def\Tr{{\text{Tr}}}

\NewDocumentCommand{\del}{e_}{%
	\nabla
	\IfValueT{#1}{%
		_{\!\!#1}
	}%
}

\begin{document}

\title{Towards the Virasoro-Shapiro Amplitude in AdS$_5\times$S$^5$}
\author{Theresa Abl, Paul Heslop and  Arthur E. Lipstein \vspace{7pt}\\ \normalsize \textit{
Department of Mathematical Sciences}\\\normalsize\textit{Durham University, Durham, DH1 3LE, United Kingdom}}
\maketitle
\begin{abstract} 
\noindent
We propose a systematic procedure for obtaining all single trace $1/2$-BPS correlators in $\mathcal{N}=4$ super Yang-Mills corresponding to the four-point tree-level amplitude for type IIB string theory in AdS$_5\times$S$^5$. The underlying idea is to compute generalised contact Witten diagrams coming from a 10d effective field theory on AdS$_5\times$S$^5$ whose coefficients are fixed by the flat space Virasoro-Shapiro amplitude  up to ambiguities related to commutators of the 10d covariant derivatives which require additional information such as localisation. We illustrate this procedure by computing stringy corrections to the supergravity prediction for all single trace $1/2$-BPS correlators up to $\mathcal{O}(\alpha'^7)$, and spell out a general algorithm for extending this to any order in $\alpha'$.   
\end{abstract}
\thispagestyle{empty}
\pagebreak
\tableofcontents

\section{Introduction}

In flat space, the four-point amplitude of closed string theory takes a very compact form known as the Virasoro-Shapiro (VS) amplitude%
\footnote{The Virasoro amplitude is the amplitude for four tachyonic scalars in bosonic string theory found  by Virasoro~\cite{Virasoro:1969me} and generalised to $n$ points by Shapiro~\cite{Shapiro:1970gy}. The tree-level four-point amplitude in IIB string theory~\cite{Green:1987sp} which we will be considering here is given by  the Virasoro amplitude multiplied by a kinematic factor and it has become the convention to refer to it as the Virasoro-Shapiro amplitude.} 
\cite{Virasoro:1969me,Green:1987sp}.  This formula encodes many essential properties of string theory such as a Regge trajectory describing massive states with arbitrarily high spin, and exponential suppression at high-energy which was one of the earliest indications that string theory could be a promising candidate for quantum gravity. Given that the effects of quantum gravity are expected to become most important in curved backgrounds like the interior of black holes and the early Universe, it is therefore very important to understand how to generalise the VS amplitude beyond the flat space limit. At present it is technically challenging to calculate string amplitudes in curved backgrounds from first principles, but progress can be made in AdS backgrounds using holographic methods. In particular IIB string theory in AdS$_5\times$S$^5$ is the best understood example due to its duality to $\mathcal{N}=4$ SYM~\cite{Maldacena:1997re}. 

This relates IIB gravity amplitudes to 
$\mathcal{N}=4$ SYM single trace%
\footnote{In fact the operators dual to supergravity are only single trace in the large $N$ limit but have multi-trace corrections at subleading order~\cite{Arutyunov:2000ima,Arutyunov:2018tvn}. These have recently been given explicitly to all orders in $N$~\cite{Aprile:2020uxk}. Here, however we work at leading order and so these multi-trace corrections will play no role.
}
$1/2$-BPS correlators. From the early days of the AdS/CFT correspondence, many direct calculations of four-point AdS amplitudes at  tree-level and in the  supergravity limit have been performed, resulting in predictions for the corresponding correlators on the CFT side~\cite{hep-th/9807097,hep-th/9903196,hep-th/9912210,hep-th/0002170,hep-th/0212116,hep-th/0301058,hep-th/0601148,0709.1365,0811.2320,1106.0630,1806.09200,1808.06788}.
Although the action for superstrings in AdS$_5\times$S$^5$ is known using the Green-Schwarz~\cite{Metsaev:1998it,Arutyunov:2009ga} and pure spinor~\cite{Berkovits:2000fe} formalisms, explicit construction of vertex operators is not fully understood so computing amplitudes beyond the supergravity approximation in this background directly from string theory remains challenging (see~\cite{Berkovits:2012ps,Azevedo:2014rva,Chandia:2017afc} for recent progress). 
On the other hand, a great deal of progress has recently been achieved on the CFT side despite the CFT being strongly coupled, using the constraints imposed by superconformal and crossing symmetry as well as the simplification of the spectrum predicted by AdS/CFT (hereby summarised as `bootstrap methods'). All tree-level single trace $1/2$-BPS correlators in the supergravity limit have been obtained in this way~\cite{Rastelli:2016nze,Rastelli:2017udc} 
and more recently string corrections have also been bootstrapped~\cite{1411.1675,1809.10670,Binder:2019jwn,Drummond:2019odu,Drummond:2020dwr,Aprile:2020luw} with groundwork laid in in~\cite{Heemskerk:2009pn,Alday:2014tsa}.
Loop corrections to four-point AdS amplitudes have also been obtained via bootstrap methods both in the supergravity limit~\cite{1706.02822,1711.02031,1711.03903,1912.01047,1912.02663} as well as string corrections~\cite{1809.10670,1812.11783,1912.07632,2008.01109}.
The more recent of these works have also made use of a hidden 10d conformal symmetry~\cite{Caron-Huot:2018kta}.

This paper can be viewed as partly going back  to  the direct calculation approach but in a hugely simplified form.  We notice that the  tree-level string corrections obtained via bootstrap methods can be obtained via  AdS$\times$S contact diagrams arising from a simple 10d  scalar effective action.
The starting point is the observation that 
if we write the flat space VS amplitude as an infinite series in $\alpha'$ (which is proportional to the square of the string length), the leading term will describe supergravity while higher order terms describe string corrections,  which can be derived from a simple effective field theory consisting of a scalar field with quartic interactions. For example, the first string correction is simply a constant proportional to $\alpha'^3$ which arises from a $\phi^4$ interaction, and the next correction is $\mathcal{O}(\alpha'^5)$ and quadratic in the Mandelstam variables so can be derived from a four-derivative interaction $(\partial \phi . \partial \phi)^2$. In this way, we can construct the four-field piece of the linearised (about flat space) effective action at all orders in $\alpha'$, fixing coefficients by comparing to the VS amplitude. This can be made more precise. All the fields of type IIB  supergravity can be described with a chiral scalar superfield, $\phi$,  in  10d $\cN=2$ superspace~\cite{Howe:1983sra}, and it is this scalar superfield that appears in the superaction. The Virasoro-Shapiro amplitude for IIB string theory is a superamplitude containing a factor $\delta^{16}(Q)$~\cite{Green:2019rhz}. Similarly the corresponding linearised effective action is a superaction and one integrates a scalar superfield (prepotential) over 16 Grassmann odd variables $\int d^{16}\theta$~\cite{Green:1999qt}. The action of four Grassman derivatives on the scalar produces the Riemann curvature and so $\phi^4$ in the effective superpotential produces the familiar $R^4$ correction to supergravity.

Remarkably,  we will find  that the interacting part of all single trace $1/2$-BPS   correlators  
can be obtained from a similar
scalar effective action describing tree-level IIB string theory on AdS$_5\times$S$^5$ (rather than a  flat) background. The resulting correlators are naturally packaged together into a 10d structure. This 10d structure is very reminiscent of and indeed was partly inspired by the 10d conformal structure of these correlators observed in~\cite{Caron-Huot:2018kta}. However, here  the 10d {\em conformal } structure is not apparent and does not play a role. We can read off some coefficients of the  AdS$\times$S  effective action directly from the flat space one, but  not all terms  can be read off in this way. Firstly, since covariant derivatives will no longer commute in general, there is the possibility of commutator terms which vanish in flat space. Furthermore it is also  possible to add terms proportional to the curvature which vanish in the flat space limit. The effective action will therefore have additional terms with unfixed coefficients. 

We do not here prove the existence of the effective field theory on AdS$_5\times $S$^5$, but justify it a posteriori by showing that it reproduces all known results for four-point correlators of single trace  $1/2$-BPS operators at orders in $\alpha'^3$ and $\alpha'^5$, which were previously obtained via bootstrap methods in~\cite{1411.1675,1809.10670,Binder:2019jwn,Drummond:2019odu,Drummond:2020dwr,Aprile:2020luw}. We also present a general algorithm for extending these predictions to arbitrarily high order in $\alpha'$ and use it to obtain new predictions at $\alpha'^6$ and $\alpha'^7$. A key technical tool that allows us to derive correlation functions from the 10d effective field theory is the use of  a natural generalisation  of contact Witten diagrams~\cite{Witten:1998qj} (which are integrals over AdS space) to integrals over the full AdS$\times$S space,  treating AdS and S on an equal footing. We are not aware of such generalised Witten diagrams directly appearing in the literature before, although similar structures  on the sphere are given in~\cite{Chen:2020ipe} where analogues of  geodesic Witten diagrams (which give conformal blocks)  on the sphere were considered. The generalised Witten diagrams involve introducing propagators connecting the $(5+5)$-dimensional bulk of AdS$_5\times$S$^5$ to a generalised notion of a boundary. Although the 5-sphere is compact, we can formally define its boundary using embedding coordinates analogous to those of AdS$_5$. This definition is physically sensible when describing $1/2$-BPS operators since it essentially encodes the condition that they are traceless and symmetric in R-symmetry indices. Expanding the 10d Witten diagrams in modes on the S$^5$ then gives a prediction for all four-point correlators of single trace $1/2$-BPS operators corresponding to a fixed order in the $\alpha'$ expansion of string theory in AdS$_5\times$S$^5$. Comparing these results to those obtained using localisation techniques~\cite{Binder:2019jwn,Chester:2020dja,Chester:2020vyz} allows us to fix some ambiguities in the effective action. 

This paper is organised as follows. In section~\ref{setup} we provide an overview of the general strategy including a general discussion of the effective action and define generalised contact diagrams in AdS$\times$S as well as their Mellin transforms. In section~\ref{alphap3} we use these techniques to compute the leading correction to $1/2$-BPS correlators which occurs at $\alpha'^3$. In section~\ref{wdiagal}  we develop an algorithm for extending these calculations to arbitrary order in $\alpha'$. Using this algorithm, we reproduce previous results at $\alpha'^5$ in section~\ref{sec:alpha5}, and obtain new predictions at $\alpha'^6$ and $\alpha'^7$ in sections~\ref{sec:alpha6} and~\ref{sec:alpha7}, respectively. We present conclusions and future directions in section~\ref{conclusion}. There are also two appendices. In appendix~\ref{intriligator}, we present more details about the parametrisation of $1/2$-BPS correlators, and in appendix~\ref{appendix:alpha7ambiguities} we list further results at $\alpha'^7$.

{\bf Note added:}  whilst completing this we were informed by the authors
of~\cite{adps} of their very impressive work obtaining higher orders in $\alpha'$ corrections on AdS$_5\times$S$^5$ via bootstrap methods, nicely complementing the results here. We thank them for coordinating the arXiv release.

\section{General setup} \label{setup}

In this section we will describe the basic ingredients that we will use in this paper. In the first subsection, we will describe our strategy for deducing an effective action from the VS amplitude in flat space and translating it to AdS$_5\times$S$^5$. In the next section, we review $1/2$-BPS correlators in $\mathcal{N}=4$ SYM, which will be the analogue of the Virasoro amplitude in AdS$_5\times$S$^5$. In the next subsection, we review the embedding space for AdS$_5$ and S$^5$ and explain how to define covariant derivatives and contact Witten diagrams in these coordinates. In the next subsection we then show how to compute contact diagrams directly in this product space using novel bulk-to-boundary propagators which are manifestly ten-dimensional. For a given order in the $\alpha'$ expansion of the Virasoro amplitude, this will allow us to compute the infinite tower of $1/2$-BPS correlators by computing Witten diagrams from a 10d effective action and expanding them in modes on the sphere. The correlators are most elegantly expressed in Mellin space, which we review in the last subsection. In particular, we find that expanding our 10d Witten diagrams in terms of spherical coordinates gives rise to a spherical analogue of the Mellin transform and implies a generalised Mellin amplitude where AdS$_5$ and S$^5$ are on equal footing. We illustrate this approach by deriving a formula for all single trace $1/2$-BPS four-point correlators in the supergravity approximation. The question of stringy corrections will be addressed in subsequent sections.

\subsection{Effective action} \label{effectiveact}

The  flat space 
Virasoro-Shapiro amplitude takes the form
\begin{align}
A_{VS}(S,T)=\frac1{STU}\frac{\Gamma(1-\frac{\alpha'S}4)\Gamma(1-\frac{\alpha'T}4)\Gamma(1-\frac{\alpha'U}4)}{\Gamma(1+\frac{\alpha'S}4)\Gamma(1+\frac{\alpha'T}4)\Gamma(1+\frac{\alpha'U}4)}\ , \qquad \qquad S+T+U=0\ ,
\end{align} 
where $S,T,U$ are the standard four-point kinematic invariants.
This has expansion
\begin{align}\label{vs}
A_{VS}(S,T)&=\frac1{STU} \exp\left({\sum_{n=1}^\infty 2\left(\frac{\alpha'}4 \right)^{2n+1} \frac{\zeta_{2n+1}}{2n+1} (S^{2n+1}{+}T^{2n+1}{+}U^{2n+1}) }\right) \notag\\	
&= \frac1{STU} +2\zeta_3(\tfrac{\alpha'}4)^3+(S^2+T^2+U^2)\zeta_5(\tfrac{\alpha'}4)^5+2 S T U (\zeta_3)^2(\tfrac{\alpha'}4)^6\notag\\
&+\tfrac12 (S^2+T^2+U^2)^2\zeta_7(\tfrac{\alpha'}4)^7 +\dots
\end{align}
Excluding the first term, which corresponds to supergravity, we can view the remaining terms as arising from a scalar effective action. From this point of view, the $\alpha'^3$ correction which gives a constant contribution to the four-point amplitude,  comes from a $\phi^4$ interaction. Higher order terms can then be obtained by applying derivatives to the  $\phi^4$ interaction corresponding to the invariants $S,T,U$. So $S=-2k_1.k_2\rightarrow 2\partial_\mu \phi\partial^\mu \phi \phi^2$, $T=-2k_1.k_3\rightarrow 2\partial_\mu \phi \phi \partial^\mu \phi \phi$ etc. 

Specifically then the VS amplitude is equivalent to the following four-field terms in an effective superpotential for supergravity linearised about flat space:
\begin{align}\label{seff}
&  V^{\text{flat}}_{VS}(\phi)=\tfrac1{2^3.4!}\Big(2\zeta_3(\tfrac{\alpha'}2)^3\phi^4+3\zeta_5(\tfrac{\alpha'}2)^5(\partial\phi.\partial\phi)^2+2 (\zeta_3)^2(\tfrac{\alpha'}2)^6(\partial\phi.\partial\phi)(\partial_\mu\partial_\nu\phi\partial^\mu\partial^\nu\phi)\notag \\
&\qquad \qquad \qquad \qquad \qquad \qquad +3 \zeta_7(\tfrac{\alpha'}2)^7 (\partial_\mu\partial_\nu\phi\partial^\mu\partial^\nu\phi)^2 +\dots\Big)\ .
\end{align} 
We now uplift the effective superpotential to an AdS$_5\times$S$^5$ background by replacing the flat derivatives with covariant AdS$\times$S derivatives. This uplift is not unique however. Firstly the covariant derivatives no longer commute with each other leading to ambiguities. Secondly there could be terms involving  lower number of derivatives, compensated by the AdS radius, $R$ which would vanish in the flat space limit. So to $\cO(\alpha'^7)$ the superpotential translates to 
\begin{align}\label{Seff}
	V^{\text{AdS$\times$S}}_{VS}(\phi)=
	\tfrac1{8.4!}\Bigg(&(\tfrac{\alpha'}2)^3A\phi^4+(\tfrac{\alpha'}2)^5 \Big(3B(\del\phi.\del\phi)^2+ 6C \nabla^{2}\nabla_{\mu}\phi\nabla^{\mu}\phi\phi^{2} \Big)\notag\\
&+(\tfrac{\alpha'}2)^6\Big(D (\nabla\phi.\nabla\phi)(\nabla_\mu\nabla_\nu\phi\nabla^\mu\nabla^\nu\phi)+6E \nabla_{\mu}\nabla^{2}\nabla_{\nu}\phi\nabla^{\mu}\nabla^{\nu}\phi\phi^{2}\Big)
	\notag \\
& +
(\tfrac{\alpha'}2)^7 \Big(6 F (\nabla_\mu\nabla_\nu\phi\nabla^\mu\nabla^\nu\phi)^2 +6 G_1\left(\nabla^\mu\nabla^\nu\nabla_\mu\nabla^\rho\nabla^\sigma\nabla_\rho\phi\right)\left(\nabla_\nu\nabla_\sigma\phi\right)\phi^2+\dots\Big)+\dots\Bigg)\ .
\end{align}
There are four more eight-derivative terms with coefficients $G_2,G_3,G_4,G_5$ whose explicit expressions are given in~\eqref{alpha7ambs} and appendix~\ref{appendix:alpha7ambiguities}.
Here, unlike in flat space, the coefficients $A,B,C,..$ themselves can have an expansion in the dimensionless parameter $\alpha'/R^2$ where $R$ is the radius of AdS (or S). So whereas in flat space $2k$-derivative terms only occur at order $\alpha'^{k+3}$, in AdS$\times$S, $2k$-derivative terms  occur at $\alpha'^{k+3}$ and all higher orders in principle. 

The zeroth order terms in the expansion of $A,B,D,F$ are then determined by the Virasoro amplitude. Specifically, then 
\begin{align}\label{coeffexpansion}
	A({\alpha'}) &=2\zeta_3+A_1\tfrac{\alpha'}{2R^2}+A_2\left(\tfrac{\alpha'}{2R^2}\right)^2+\dots\notag \\
	B({\alpha'}) &=\zeta_5+B_1\tfrac{\alpha'}{2R^2}+\dots\notag \\
	C({\alpha'}) &=C_0+C_1\tfrac{\alpha'}{2R^2}+\dots\notag \\
	D({\alpha'}) &=2(\zeta_3)^2+D_1\tfrac{\alpha'}{2R^2}+\dots\notag \\
	E({\alpha'}) &=E_0+E_1\tfrac{\alpha'}{2R^2}+\dots\notag \\
	F({\alpha'}) &=\tfrac{1}2 \zeta_7+F_1\tfrac{\alpha'}{2R^2}+\dots\notag \\
	G_i({\alpha'}) &=G_{i;0}+G_{i;1}\tfrac{\alpha'}{2R^2}+\dots \qquad \qquad \text{for $i=1,2,3,4,5$}\
\end{align}
For simplicity, we will set $R=1$ from now on throughout this paper, but it will be understood that these higher order terms vanish in the flat space limit. 

Computing 10d Witten diagrams using novel generalised bulk-to-boundary propagators and expanding them in terms of $S^5$ coordinates will give all single trace $1/2$-BPS four-point correlators in $\mathcal{N}=4$ SYM described by tree-level string theory in AdS$_5 \times$S$^5$. We introduce all of these things in the following subsections. 

\subsection{$1/2$-BPS correlators}

In $\cN=4$ SYM there are six real scalars transforming in the adjoint rep of $SU(N)$ and the fundamental rep of $SO(6)$,  $\phi_{YM}^I(X)$. Here we view the 4d Minkowski space via null 6d embedding coordinates $X^A$ with $X.X=0$ manifesting the conformal $SO(2,4)$ symmetry.  We also project with a null 6d coordinate $Y_I,\  Y.Y=0$ to obtain  $\phi_{YM}(X,Y)=\phi_{YM}^I(X)Y_I$ Manifesting the internal $SO(6)$ symmetry. Then the single trace $1/2$-BPS operators are defined as
\begin{equation}
	\cO_p(X,Y)=\frac{1}{p N^{p/2}} \Tr(\phi_{YM}^p). 
\end{equation}
Note that we normalise the operators with an additional factor of $1/\sqrt{p}$ compared to the normalisation giving a normalised two-point function, first derived in~\cite{Lee:1998bxa}. This normalisation is inspired by the ten-dimensional conformal symmetry of~\cite{Caron-Huot:2018kta}. 

It is then useful to collect together the  four-point functions of all single trace $1/2$-BPS operators $\cO_p(X,Y)$ into a single object $\langle \cO\cO\cO\cO \rangle$ as follows
\begin{align}\label{oooo}
 \langle \cO\cO\cO\cO\rangle &= \sum_{p,q,r,s=2}^\infty {\langle \cO_p \cO_q \cO_r \cO_s\rangle_{\text{int}}}\ ,
 \end{align}
 where $\langle \cO_p \cO_q \cO_r \cO_s\rangle_{\text{int}}$ represents the interacting part of the correlator, which always contains a particular factor $I(X_i,Y_i)$ due to superconformal symmetry~\cite{Eden:2000bk} which we thus divide out
 \begin{align}\label{4intdef}
 	\langle \cO_p \cO_q \cO_r \cO_s\rangle_{\text{int}} = \frac{\langle \cO_p \cO_q \cO_r \cO_s\rangle-\langle \cO_p \cO_q \cO_r \cO_s\rangle_{\text{free}}}{I(X_i,Y_i)}\ .
 \end{align}
 From now on we will usually drop the explicit `int' subscript at the end of the correlators. 
 
 Here $I$ is a polynomial in $X_i$ and $Y_i$ which is a common factor of all interacting $1/2$-BPS four-point functions~\cite{Eden:2000bk}. 
 It is the counterpart of the  $\delta^{16}(Q)$ factor of flat space superamplitudes~\cite{Green:2020eyj}.
 We give its explicit form in appendix~\ref{intriligator}.

\subsection{Generalised contact Witten diagrams}

In this subsection we will first review standard AdS contact Witten diagrams. Then we define  analogous objects on the sphere (following similar ideas in~\cite{Chen:2020ipe}) and
finally we
introduce a generalisation of Witten diagrams using bulk-to-boundary propagators which are intrinsically ten-dimensional and treat AdS and S on equal footing. This will have a big pay-off since we will obtain the whole tower of $1/2$-BPS correlators by expanding the Witten diagrams in spherical coordinates.

The Witten diagrams are most conveniently expressed using embedding coordinates for both AdS$_{d+1}$ and S$^{d+1}$:
\begin{align}
&\hat{X}^{2}=-\left(\hat{X}^{-1}\right)^{2}-\left(\hat{X}^{0}\right)^{2}+\sum_{i=1}^{d}\left(\hat{X}^{i}\right)^{2}=-1\ ,\notag\\
&\hat{Y}^{2}=\sum_{i=-1}^{d}\left(\hat{Y}^{i}\right)^{2}=1\ .
\end{align}
In the present context, $d=4$. 
In terms of these coordinates, covariant derivatives can be defined
using projection tensors
\begin{align}\label{pdef}
\mathcal{P}_{A}^{B}=\delta_{A}^{B}+\hat{X}_{A}\hat{X}^{B}\ ,\qquad \mathcal{P}_{I}^{J}=\delta_{I}^{J}-\hat{Y}_{I}\hat{Y}^{J}\ ,
\end{align}
which satisfy the useful identities
\begin{align}\label{useful}
\mathcal{P}_{A}^{B}\hat{X}^{A}&=0\ , \qquad &\mathcal{P}_{I}^{J}\hat{Y}^{J}&=0\ ,\notag\\
\mathcal{P}_{A}^{B}\mathcal{P}_{B}^{C}\, &=\mathcal{P}_{A}^{C}\ ,&\mathcal{P}_{I}^{J}\mathcal{P}_{J}^{K}&=\mathcal{P}_{I}^{K}\ .
\end{align}
In particular, the covariant derivative of a tensor is given by~\cite{Penedones:2010ue,Sleight:2016hyl}
\begin{equation}
\nabla_{A}\rm{T}_{A_{1}...A_{N}}=\mathcal{P}_{A}^{C}\mathcal{P}_{A_{1}}^{C_{1}}...\mathcal{P}_{A_{N}}^{C_{N}}\partial_{C}\left(\mathcal{P}_{C_{1}}^{E_{1}}...\mathcal{P}_{C_{N}}^{E_{N}}\rm{T}_{E_{1}...E_{N}}\right)\ .\label{eq:covariantderv}
\end{equation}
As an application, let's consider two transverse tensors $\rm{T}$ and
$\rm{U}$ of rank $N+1$ and $N$, respectively. Using the chain rule,
we see that
\begin{equation}
\rm{T}^{BA_{1}...A_{N}}\nabla_{B}\rm{U}_{A_{1}...A_{N}}=-\nabla_{B}\rm{T}^{BA_{1}...A_{N}}\rm{U}_{A_{1}...A_{N}}+...\label{eq:ibp}
\end{equation}
where the ellipsis denote
\begin{equation}
\partial_{C}\left(\rm{T}_{2}^{BA_{1}...A_{N}}\mathcal{P}_{B}^{C}\rm{U}_{A_{1}...A_{N}}\right)-\rm{T}^{BA_{1}...A_{N}}\partial_{C}\left(\mathcal{P}_{B}^{C}\mathcal{P}_{A_{1}}^{C_{1}}...\mathcal{P}_{A_{N}}^{C_{N}}\right)\rm{U}_{CC_{1}...C_{N}}\ .
\end{equation}
When we act on projection tensors with derivatives, this gives terms which vanish when contracted
with the transverse tensors, so the second term vanishes. Since the
first term is a total derivative, \eqref{eq:ibp} implies that Lagrangians
written in embedding coordinates enjoy the same equivalence relations
as flat space Lagrangians under integration by parts. The above discussion of covariant derivatives can equally be applied to the sphere case by simply sending $A,B,C$ indices to $I,J,K$ etc.

Now we first recall the standard AdS contact Witten diagrams in embedding space. These are defined as integrals over AdS${}_{d+1}$  of products of bulk-to-boundary propagators \footnote{Properly normalised to yield a delta function at the boundary, the bulk-to-boundary propagator will include a normalisation $\mathcal{C}_{\Delta} =\frac{\Gamma(\Delta)}{2{\pi}^{d/2} \Gamma(\Delta-d/2+1)}$
\cite{Klebanov:1999tb,Penedones:2010ue}. These are normally omitted from the definition of the contact diagrams or $D$ functions and we do so here. We will also  later absorb factors of $\mathcal{C}_{\Delta_i}$ into the definition of the Mellin amplitude.},
\begin{equation}
G(\hat X,X_i)=\frac{\mathcal{C}_{\Delta_i}}{(-2\hat X.X_i)^{\Delta_i}}\ ,
\end{equation}
which at four points then yields:%
\begin{align}\label{adscont}
	D^{(d)}_{\Delta_1\Delta_2\Delta_3\Delta_4}(X_i) = \frac1{(-2)^{2\Sigma_{\Delta}}}\int_\text{AdS}  \frac{d^{d+1} \hat X}{(\hat X.X_1)^{\Delta_1}(\hat X.X_2)^{\Delta_2}(\hat X.X_3)^{\Delta_3}(\hat X.X_4)^{\Delta_4}}\ ,
\end{align} 
where $\Sigma_\Delta=(\Delta_1{+}\Delta_2{+}\Delta_3{+}\Delta_4)/2$. The powers of minus 2 can be absorbed into the propagators, as  $(-2\hat{X}.X_i)$ but for notational simplicity we pull them out. 
 These $D$ functions have the following form in Mellin space~\cite{Penedones:2010ue} 
\begin{align}\label{Ddef}
	D^{(d)}_{\Delta_1\Delta_2\Delta_3\Delta_4}(X_i)= \cN^{\text{AdS}_{d+1}}_{\Delta_i} \times
	\int \frac{d\delta_{ij}}{(2 \pi i)^2}  {\prod_{i<j}\frac{\Gamma(\delta_{ij})}{(X_{i}.X_{j})^{\delta_{ij}} }   }  \qquad \qquad \sum_i \delta_{ij}=\Delta_j \ , 
\end{align}
where the normalisation is given by
\begin{align}\label{norm}
    \cN^{\text{AdS}_{d+1}}_{\Delta_i}=\frac{\tfrac12 \pi^{d/2} \Gamma(\Sigma_\Delta-d/2)}{(-2)^{\Sigma_\Delta}\prod_i\Gamma(\Delta_i)}\ .
\end{align}
For later use we define normalised $D$ functions without the factor $\cN$ as
\begin{align}\label{normalisedD}
    D_{\Delta_i}(X_i)=\cN^{-1}D^{(d)}_{\Delta_i}(X_i)\ .
\end{align}
Note that the normalised $D$ functions are independent of the spacetime dimension $d$ as can be seen from~\eqref{Ddef} and they are distinguished by the presence or not of the superscript $(d)$. 

We can also consider direct analogues of these contact diagrams on the sphere.  Bulk-to-boundary propagators on the sphere were introduced  in~\cite{Chen:2020ipe} $G(\hat Y,Y_i)\propto (-2\hat Y.Y_i)^{p_i}$ and 
in this context it is then very natural to introduce functions  $B_{p_1p_2p_3p_4}(Y_i)$,  spherical analogues of the  contact Witten diagrams $D_{\Delta_1\Delta_2\Delta_3\Delta_4}$ as:
\begin{align}\label{scont}
	B_{p_1p_2p_3p_4}(Y_1,Y_2,Y_3,Y_4)= (-2)^{2\Sigma_p}\int_{\text{S}} {d^{d+1} \hat Y}   (\hat Y.Y_1)^{p_1}(\hat Y.Y_2)^{p_2}(\hat Y.Y_3)^{p_3}(\hat Y.Y_4)^{p_4} \ ,
\end{align}
where $\Sigma_p=(p_1{+}p_2{+}p_3{+}p_4)/2$.
Even though the sphere is compact, we can formally define a boundary when describing $1/2$-BPS operators in $\mathcal{N}=4$ SYM since the condition $Y.Y=0$ simply encodes tracelessness of the R-symmetry indices.
The $B$ functions are polynomials in the $Y_i$ and can be explicitly evaluated purely combinatorially, following similar techniques to those found in the appendix of~\cite{Chen:2020ipe} (where the two- and three-point analogues were obtained):
\begin{align}\label{bsum}
	B_{p_1p_2p_3p_4}(Y_i) =   \cN^{\text{S}^{d+1}}_{p_i}\sum_{\{d_{ij}\}}  \prod_{i<j} \frac{(Y_{i}.Y_{j})^{d_{ij}}}{\Gamma(d_{ij}+1)} \qquad \qquad \sum_i d_{ij}=p_j\ ,
\end{align}
where 
\begin{align}
    \cN^{\text{S}^{d+1}}_{p_i}={2.2^{\Sigma_p}}\frac{\pi^{d/2+1}\prod_i\Gamma(p_i{+}1)}{\Gamma(\Sigma_p{+}d/2{+}1)} \ .
\end{align}
In~\eqref{bsum} the sum is over all sets of numbers $d_{ij}=d_{ji}$ such that
\begin{align}
	\Big\{(d_{12},d_{13},d_{14},d_{23},d_{24},d_{34}): 0\leq d_{ij}=d_{ji}, \quad d_{ii}=0,\quad  \sum_{i=1}^4 d_{ij} =p_j  \Big\}\ .
\end{align}
	These constraints on $d_{ij}$  leave just two free parameters.
Note the close similarity this explicit expansion of the $B$ functions~\eqref{bsum} has with the Mellin transform of the AdS contact terms~\eqref{Ddef}. We can thus view the expansion parameters $d_{ij}$ as analogues of the Mellin variables $\delta_{ij}$.

It is now natural to combine the above AdS and S bulk-to-boundary propagators into one 10d object, which we refer to as a generalised bulk-to-boundary propagator in AdS$\times$S:
\begin{align}\label{bbprop}
	G(\hat X,\hat Y ; X,Y)=\left(-2\hat{X}. X-2\hat{Y}. Y\right)^{-\Delta}\ ,
\end{align}
where $X$ and $Y$ satisfy 
\begin{equation}
X^{2}=Y^{2}=0\ .
\end{equation}
Using the definition in~\eqref{eq:covariantderv}, we see that
\begin{align}
\nabla^{2}G=\left(\nabla_{\hat{X}}^{2}+\nabla_{\hat{Y}}^{2}\right)G=\Delta(\Delta-d)\left((-2\hat{X}. X)^{2}-(-2\hat{Y}. Y)^{2}\right)\left(-2\hat{X}. X-2\hat{Y}. Y\right)^{-\Delta-2}\ .
\end{align}
Hence, the propagator obeys massless equations of motion when $d=\Delta$:
\begin{equation}
	\nabla^{2}G=0\ ,\label{eom}
\end{equation}
which will become important in the next section. Whereas $X$ describes the boundary of AdS, $Y$ is not a boundary point since the sphere is compact. 

As mentioned in the introduction, we will derive predictions for four-point correlators of $1/2$-BPS operators from an effective action by computing analogues of Witten diagrams directly in the product geometry AdS$\times$S. For now we will just develop some general properties of AdS${}_{d+1}\times$S$^{d+1}$ contact Witten diagrams which are defined simply as
\footnote{We keep $d$ and $\Delta_i$ general here but we will be focussing on the case $\Delta_i=d=4$ later.}
\begin{align}\label{adsxscont}
	D^{\text{AdS}_{d+1}\times \text{S}^{d+1}}_{\Delta_1\Delta_2\Delta_3\Delta_4}(X_i,Y_i) = \frac1{(-2)^{2\Sigma_{\Delta}}}\int_{\text{AdS}\times\text{S}}  \frac{d^{d+1} \hat X d^{d+1} \hat Y}{(P_1+Q_1)^{\Delta_1}(P_2+Q_2)^{\Delta_2}(P_3+Q_3)^{\Delta_3}(P_4+Q_4)^{\Delta_4}}\ ,
\end{align}
where we introduce the shorthand
\begin{align}
	P_i=\hat X.X_i\ , \qquad Q_i=\hat Y.Y_i\ .
\end{align}
It is then straightforward to expand this AdS$\times$S contact diagram out into an infinite number of standard AdS contact diagrams multiplied by sphere analogues. In particular, using
\begin{align}\label{exp}
	\frac1{(P+Q)^{\Delta}}=\sum_{p=0}^\infty(-1)^p \frac{(p+1)_{\Delta-1}}{\Gamma(\Delta)}\frac{Q^p}{P^{p+\Delta}}
\end{align}
four times and then inserting~\eqref{adscont} and~\eqref{scont} gives the expansion:%

\begin{align}\label{DinDB}
	D^{\text{AdS}\times\text{S}}_{\Delta_1\Delta_2\Delta_3\Delta_4}(X_i,Y_i) 
	= \sum_{p_i=0}^\infty 
	\prod_{i=1}^4(-1)^{p_i}\frac{(p_i+1)_{\Delta_i-1}}{\Gamma(\Delta_i)} 
	D^{(d)}_{p_1+\Delta_1,p_2+\Delta_2,p_3+\Delta_3,p_4+\Delta_4}(X_i) 
        B_{p_1p_2p_3p_4}(Y_i) \ .
\end{align}

\subsection{AdS$\times$S contact diagrams in Mellin space}

Inserting the expression for the AdS contact term, $D$,  as a Mellin integral~\eqref{Ddef} and the sphere analogue $B$ as an expansion~\eqref{bsum} into the expression for the AdS$\times$S contact term~\eqref{DinDB}, we get after some simplifications a Mellin representation for the AdS$\times$S contact term:
\begin{align}\label{mellin10d}
	&D^{\text{AdS}_{d+1}\times\text{S}^{d+1}}_{\Delta_1\Delta_2\Delta_3\Delta_4}(X_i,Y_i) \notag \\
	&=
\frac{	\pi^{d+1}}{(-2)^{\Sigma_\Delta}\prod_i\Gamma(\Delta_i)}\times \sum_{p_i=0}^\infty(-1)^{\Sigma_p}\int \frac{d\delta_{ij}}{(2 \pi i)^2} \sum_{(d_{ij})} \, \left(\prod_{i<j}\frac{(Y_i.Y_j)^{d_{ij}}}{(X_i.X_j)^{\delta_{ij}}}\frac{\Gamma(\delta_{ij})}{\Gamma(d_{ij}+1)}\right)\times (\Sigma_p{+}d/2{+}1)_{\Sigma_\Delta-d-1}\ ,\notag\\
& \qquad \qquad \text{where} \qquad \sum_{i\neq j} \delta_{ij}=p_j+\Delta_j \qquad \qquad \qquad 	\sum_{i\neq j} d_{ij}=p_j\ .
\end{align}

We thus define the AdS$_{d+1}\times$S$^{d+1}$ Mellin amplitude, $\cM_{\Delta_i}[f](\delta_{ij},d_{ij})$, for any such four-point expression, $f$, via a similar expression
\begin{align}
\label{mellinampdef}
    &f(X_i,Y_i)	\notag\\
    &=\frac1{4!}
\frac{	\pi^{d+1}}{(-2)^{\Sigma_\Delta}}\left(\prod_i\frac{\mathcal{C}_{\Delta_i}}{\Gamma(\Delta_i)}\right)\times \sum_{p_i=0}^\infty(-1)^{\Sigma_p}\int \frac{d\delta_{ij}}{(2 \pi i)^2} \sum_{(d_{ij})} \, \left(\prod_{i<j}\frac{(Y_i.Y_j)^{d_{ij}}}{(X_i.X_j)^{\delta_{ij}}}\frac{\Gamma(\delta_{ij})}{\Gamma(d_{ij}+1)}\right)\times \cM_{\Delta_i}[f]\ ,\notag\\
& \qquad \qquad \text{where} \qquad \sum_{i\neq j} \delta_{ij}=p_j+\Delta_j \qquad \qquad \qquad 	\sum_{i\neq j} d_{ij}=p_j\ .
\end{align}
Thus the Mellin amplitude  of an AdS$\times$S contact diagram is not in general a constant as for the AdS case, but rather a Pochhammer: 
\begin{align}\label{mellinampcont}
\frac1{4!} \left( {\prod_i\mathcal{C}_{\Delta_i}} \right)\times  D^{\text{AdS}_{d+1}\times \text{S}^{d+1}}_{\Delta_1\Delta_2\Delta_3\Delta_4}(X_i,Y_i) \qquad \leftrightarrow \qquad  \cM_{\Delta_i}(\delta_{ij},d_{ij}) = { (\Sigma_p{+}d/2{+}1)_{\Sigma_\Delta-d-1}}\ .
    \end{align}

\subsection{Relation between contact diagrams in AdS$\times$S and AdS}
 
Although we will not use this fact in the rest of the paper it is worth pointing out here an intriguing relation between the AdS$\times$S contact diagrams and the better known standard AdS contact diagrams.
This relation can be seen by comparing their respective Mellin transforms~\eqref{mellin10d} and~\eqref{Ddef}.

First consider the special case $\Sigma_\Delta=d+1$. In this case the final Pochhammer in~\eqref{mellin10d} is absent and the Mellin transform becomes proportional to
\begin{align}\label{sumup}
 &\sum_{p_i=0}^\infty (-1)^{\Sigma_p}	\int \frac{d\delta_{ij}}{(2 \pi i)^2} \sum_{(d_{ij})} \, \left(\prod_{i<j}\frac{(Y_i.Y_j)^{d_{ij}}}{(X_i.X_j)^{\delta_{ij}}}\frac{\Gamma(\delta_{ij})}{\Gamma(d_{ij}+1)}\right) = \int \frac{d{\pmb \delta}_{ij}}{(2 \pi i)^2} \, \left(\prod_{i<j}\frac{\Gamma({\pmb \delta}_{ij})}{(X_i.X_j+Y_i.Y_j)^{{\pmb \delta}_{ij}}}\right)\ ,\notag \\
&\qquad \qquad \text{where} \qquad
	\sum_{i\neq j} {\pmb \delta}_{ij}=\Delta_j\ ,
\end{align}
where the equality is obtained  by performing the sums over  $p_{i}$ and then changing variables from $\delta_{ij} \rightarrow  {\pmb \delta}_{ij} =\delta_{ij} -d_{ij}$. 

Comparing this with the Mellin transform of the AdS contact term~\eqref{Ddef} we see that this is proportional to a $D$ function with $X_i.X_j \rightarrow X_i.X_j+Y_i.Y_j$. In other words it is proportional to a pure AdS contact term with embedding coordinates $X_i^\mu=(X_i^A,Y_i^I)$, corresponding to a $(2d+2)$-dimensional bulk. 
More precisely we have the relation \footnote{We here compare with the dimension independent, normalised $D$ function~\eqref{normalisedD}, since in $2d+2$ dimensions the $D^{(d)}$ function itself diverges when  $\Sigma_\Delta=d+1$ due to the 
$\Gamma$ in the numerator of~\eqref{norm}.} 
\begin{align}\label{relateDsspecial}
	D^{\text{AdS}_{d+1}\times\text{S}^{d+1}}_{\Delta_1\Delta_2\Delta_3\Delta_4}(X_i,Y_i)=
\frac{	\pi^{d+1}}{(-2)^{\Sigma_\Delta}\prod_i\Gamma(\Delta_i)}\times  D_{\Delta_1\Delta_2\Delta_3\Delta_4}(X_i,Y_i)\qquad  \Sigma_\Delta=d+1\ .
\end{align}
Note that this case $\Sigma_\Delta= d+1$ corresponds precisely to the case of  a dimensionless contact term in the flat space limit, $\int d^{2d+2}x \phi_{\Delta_1} ...\phi_{\Delta_4}$. The above  relation~\eqref{relateDsspecial} is an example of the enhanced higher dimensional conformal symmetry observed in~\cite{Caron-Huot:2018kta}. We will look at this explicitly for $\cN=4$ SYM in the next subsection.

Now let us modify the above discussion for the case with $\Sigma_\Delta\neq d+1$. Here the direct relation between AdS$\times$S and AdS contact terms is spoiled by the presence of the Pochhammer at the end of~\eqref{mellin10d} which depends on $\Sigma_p$ that we are summing over. A simple way of reproducing this Pochhammer whilst still having a summed up formula is then to rescale all the $Y$ variables and differentiate. 
Concretely, we can write
\begin{align}\label{relateDs}
	D^{\text{AdS}_{d+1}\times\text{S}^{d+1}}_{\Delta_1\Delta_2\Delta_3\Delta_4}(X_i,\sqrt{r}Y_i)=
\frac2{\Gamma(\Sigma_\Delta-d-1)}	\frac 1{r^{d/2}}\left(\frac d {dr}\right)^{\Sigma_\Delta-d-1} r^{\Sigma_\Delta-d/2-1}D^{(2d+2)}_{\Delta_1\Delta_2\Delta_3\Delta_4}(X_i,\sqrt{r}Y_i)\ , 
\end{align}
where the $D$ function is for a $(2d+2)$-dimensional bulk.

\subsection{Tree-level supergravity}

While the main focus of this paper is obtaining tree-level string corrections to $\mathcal{N}=4$ SYM correlators from an effective action involving massless scalars in 10d, it is interesting to first look at the tree-level supergravity prediction following the approach described in the previous subsection. While we do not expect this to arise from an effective superpotential, all single trace $1/2$-BPS correlators were shown in~\cite{Caron-Huot:2018kta} to possess a 10d conformal structure and in particular can be obtained by expanding out $ D_{2422}$. Now for $D_{2422}$ we have $\Sigma_\Delta =5=d+1$ so this is  a case where the AdS contact term and the AdS$\times$S contact terms agree, so ~\eqref{relateDsspecial} applies. The tree-level SUGRA result can be written~\cite{Caron-Huot:2018kta}
\begin{align}
	\langle \cO \cO \cO\cO \rangle_{\text{sugra}} \propto\frac{1}{(X_1.X_3{+}Y_1.Y_3)}\frac{1}{(X_1.X_4{+}Y_1.Y_4)}\frac{1}{(X_3.X_4{+}Y_3.Y_4)}D_{2422}(X_i,Y_i)\, .
\end{align} 
Inserting the Mellin representation of $D_{2422}$~\eqref{Ddef} and changing variables $\delta_{ij}\rightarrow \delta_{ij}-1$ for $i,j=1,3,4$ and $\delta_{ij}$ unchanged otherwise,   this can be written in the form~\eqref{mellinampdef} with $\Delta_i=4$ with the  Mellin amplitude
\begin{align}\cM_{\text{sugra}}\propto\frac1{({\pmb \delta}_{13}{-}1)({\pmb \delta}_{14}{-}1)({\pmb \delta}_{34}{-}1)}=\frac1{(\delta_{13}{-}d_{13}{-}1)(\delta_{14}{-}d_{14}{-}1)(\delta_{34}{-}d_{34}{-}1)}\ .
\end{align}

\section{$\alpha'^3$ corrections} \label{alphap3}

Having outlined the general procedure for computing stringy corrections to four-point $1/2$-BPS correlators in $\mathcal{N}=4$ SYM using an effective action in AdS$_5\times$S$^5$, we will now illustrate how it works for the first correction to the supergravity prediction, which occurs at order $\alpha'^3$.

In particular, the first term of the effective action~\eqref{Seff} is just a $\phi^4$ interaction:%
\begin{align}
 S_{\alpha'^3} = \frac1{8.4!}\left(\frac {\alpha'}2\right)^3 \times 2\zeta_3\times \int_{\text{AdS}\times\text{S}} {d^{5} {\hat X}}{d^{5} {\hat Y}} \phi(\hat X,\hat Y)^4\ .
\end{align}
To obtain the corresponding CFT correlators we mimic the standard AdS/CFT procedure for obtaining correlators from AdS, but in a fully 10d covariant way, including the sphere manifestly. 
Using the generalised bulk-to-boundary propagators in~\eqref{bbprop} we obtain the  
AdS$\times$S Witten diagram for this contact interaction, yielding the following proposal for the $\alpha'^3$ corrections to the correlators:
\begin{align}\label{apcubed}
\langle\cO\cO\cO\cO \rangle|_{\alpha'^3} &= \frac1{8.4!}\left(\frac {\alpha'}2\right)^3 \times 2\zeta_3\times	 \frac{(\mathcal{C}_4)^4}{(-2)^{16}} \int_{\text{AdS}\times\text{S}}  
\frac{{d^{5} {\hat X}}{d^{5} {\hat Y}}}{(P_1+Q_1)^4(P_2+Q_2)^4(P_3+Q_3)^4(P_4+Q_4)^4}\notag \\
&=\frac{1}{8.4!}\left(\frac {\alpha'}2\right)^3 (\mathcal{C}_4)^4\times2\zeta_3\times D_{4444}^{\text{AdS}_5\times\text{S}^5}\ .
\end{align}

We can now extract any specific $1/2$-BPS correlator from~\eqref{apcubed} by expanding to the appropriate power in $Y_i$ (see~\eqref{oooo}). 
 First note that the 10d bulk-to-boundary propagator Taylor expands as
\begin{align}\label{expDel4}
	(P_i+Q_i)^{-4} = \sum_{p=2}^\infty (-1)^p\frac{(p-1)_3}6(P_i)^{-p-2} (-Q_i)^{p-2}\ .
\end{align}
So the individual correlators are given by%
\footnote{This is~\eqref{DinDB} with $\Delta_i=d=4$ and with $p_i\rightarrow p_i-2$ to account for the fact that the lowest correlator is labelled with $p_i=2$ rather than $p_i=0$. We do not need to worry about the minus signs  in the factors $(-1)^p$  in~\eqref{expDel4} since $B_{p_1p_2p_3p_4}=0$ if $p_1+p_2+p_3+p_4$ is odd .}:
\begin{align}\label{ap3}
 &\langle \cO_{p_1} \cO_{p_2} \cO_{p_3} \cO_{p_4} \rangle|_{\alpha'^3}\notag\\ &= \frac1{8.4!}\left(\frac {\alpha'}2\right)^3 \times 2\zeta_3\times \frac{(\mathcal{C}_4)^4}{(-2)^{16}}\prod_i\frac{(p_i-1)_3}{3!}    	\int_{\text{AdS}_5} {d^5\hat X} \prod_i\frac1{(P_i)^{p_i+2}} \times\int_{\text{S}^5} {d^5\hat Y} \prod_i(Q_i)^{p_i-2}\notag \\
	&=  \frac1{8.4!}\left(\frac {\alpha'}2\right)^3 (\mathcal{C}_4)^4\times 2\zeta_3 \times \left(\prod_i\frac{(p_i-1)_3}{3!} \right)   D^{(4)}_{p_1+2,p_2+2,p_3+2,p_4+2}(X_i) \times B_{p_1-2,p_2-2,p_3-2,p_4-2}(Y_i)\ .
\end{align} 

To see what it looks like in Mellin space we plug the Mellin transform of $D$~\eqref{Ddef} and the expansion of $B$~\eqref{bsum} into this expression (or just use~\eqref{mellin10d} ) giving the Mellin amplitude (defined in~\eqref{mellinampdef})
\begin{equation}
\mathcal{M}_{\alpha'^3}=  \tfrac18 \left(\tfrac{\alpha'}2\right)^3 \times 2\zeta_3\times (\Sigma_p{-}1)_{3}\ .
\label{melap3}
\end{equation}
 This correctly reproduces the results of~\cite{Goncalves:2014ffa,Alday:2018pdi,Drummond:2019odu} for the Mellin amplitude of $1/2$-BPS correlators at this order.

\section{Algorithm for computing general $\alpha'$ corrections} \label{wdiagal}

At higher orders in $\alpha'$ the effective action~\eqref{Seff} has terms with covariant derivatives acting on the scalar field. Thus before proceeding  
we describe an efficient way to evaluate generalised contact diagrams with derivatives in AdS$\times$S in position space. Then we present a general formula for converting them to Mellin space.

\subsection{Generalised Witten diagrams}

Computing the action of the covariant derivatives  becomes quickly quite complicated and so it is useful to develop an algorithm to do this automatically. We will motivate the algorithm by building up from simple cases. First we consider the application of multiple covariant derivatives at a single point in AdS.
From~\eqref{eq:covariantderv} this is given recursively as
\begin{align}
	\del_A \del_B ... \del_C \phi=\cP_A^{A'}\cP_B^{B'}...\cP_C^{C'} \partial_{A'} \left(\del_{B'} ... \del_{C'} \phi \right)\ ,\qquad \qquad \del_A \phi = \cP_A^{A'}\partial_{A'}\phi\ .
\end{align} 
So the application of two covariant derivatives gives
\begin{align}
	\del_B \del_A \phi = \cP_B^{B'}\cP_A^{A'}\partial_{B'}\cP_{A'}^{A''}\partial_{A''} \phi=\cP_B^{B'}\cP_A^{A'}\partial_{B'}\partial_{A'} \phi + \cP_{BA}\hat X.\partial\phi\ .
\end{align} 
The first term arises from the partial derivative $\partial_{A'}$ being commuted through $\cP_{B'}^{B''}$ whereas the second term arises from the partial derivative hitting $\cP_{B'}^{B''}$. To arrive at this form one then uses the definition of $\cP$ given in~\eqref{pdef} as well as the useful 
 formulae~\eqref{useful}. We denote this result graphically as
 \begin{align}\label{del2}
   	\raisebox{0.35cm}{$	\del_B \del_A  = \quad  $}\begin{tikzpicture}
 	\coordinate[dot,label=left:{$\scriptstyle A$}]   (A)     at   (0,0);
 	\coordinate[dot,label=left:{$\scriptstyle B$}]   (B)     at   (0,1/2);
 	\end{tikzpicture}
  	\raisebox{0.35cm}{$\quad +\quad $}
\begin{tikzpicture}
\coordinate[dot,label=left:{$\scriptstyle A$}]   (A)     at   (0,0);
\coordinate[dot,label=left:{$\scriptstyle B$}]   (B)     at   (0,1/2);
\draw[thick](A)--(B);
\end{tikzpicture} \ , \end{align}
 where each vertex corresponds to an index ordered vertically such that the bottom one is the index of the first derivative to act. An isolated vertex at position $A$ denotes $(\cP.\partial)_A$ (with the understanding that the derivative has been commuted all the way to the right)  whereas  an edge between vertices $A$ and $B$ denotes $\cP_{AB} \hat X.\partial$.

 Now consider three covariant derivatives. Here we obtain 
 \begin{align}\label{del3}
 \del_C \del_B \del_A \phi&= \cP_C^{C'}\cP_B^{B'}\cP_A^{A'}\partial_C \left( \cP_B^{B'}\cP_A^{A'}\partial_{B'}\partial_{A'} + \cP_{BA}\hat X.\partial\right)\phi\notag \\
 &=\left(\cP_C^{C'}\cP_B^{B'}\cP_A^{A'}\partial_{C'}\partial_{B'}\partial_{A'} + \cP_{CA}\hat X.\partial \partial_B+\cP_{CB}\hat X.\partial \partial_A+\cP_{BA}\hat X.\partial \partial_C+\cP_{BA} \partial_C\right)\phi\notag\\
 & 
 \raisebox{0.75cm}{$ \hskip.15cm = \hskip 2.55cm $}\begin{tikzpicture}[scale=1.50]
 \coordinate[dot,label=left:{$\scriptstyle A$}]   (A)     at   (0,0);
 \coordinate[dot,label=left:{$\scriptstyle B$}]   (B)     at   (0,1/2);
 \coordinate[dot,label=left:{$\scriptstyle C$}]   (C)     at   (0,1);
 \end{tikzpicture}
 \raisebox{0.75cm}{$\hskip.8cm+ \hskip .7cm $}
 \begin{tikzpicture}[scale=1.50]
 \coordinate[dot,label=left:{$\scriptstyle A$}]   (A)     at   (0,0);
 \coordinate[dot,label=left:{$\scriptstyle B$}]   (B)     at   (0,1/2);
 \coordinate[dot,label=left:{$\scriptstyle C$}]   (C)     at   (0,1);
 \draw[thick](A)..controls (.15,1/2)..(C);
 \end{tikzpicture} 
 \raisebox{0.75cm}{$\hskip.7cm+ \hskip .7cm $}
 \begin{tikzpicture}[scale=1.50]
 \coordinate[dot,label=left:{$\scriptstyle A$}]   (A)     at   (0,0);
 \coordinate[dot,label=left:{$\scriptstyle B$}]   (B)     at   (0,1/2);
 \coordinate[dot,label=left:{$\scriptstyle C$}]   (C)     at   (0,1);
 \draw[thick](B)--(C);
 \end{tikzpicture} 
 \raisebox{0.75cm}{$\hskip.8cm+ \hskip .7cm$}
 \begin{tikzpicture}[scale=1.50]
 \coordinate[dot,label=left:{$\scriptstyle A$}]   (A)     at   (0,0);
 \coordinate[dot,label=left:{$\scriptstyle B$}]   (B)     at   (0,1/2);
 \coordinate[dot,label=left:{$\scriptstyle C$}]   (C)     at   (0,1);
 \draw[thick](A)--(B);
 \end{tikzpicture} 
 \raisebox{0.75cm}{$\hskip.8cm+ \hskip .3cm $}
 \begin{tikzpicture}[scale=1.50]
 \coordinate[dot,label=left:{$\scriptstyle A$}]   (A)     at   (0,0);
 \coordinate[dot,label=left:{$\scriptstyle B$}]   (B)     at   (0,1/2);
 \coordinate[dot,label=left:{$\scriptstyle C$}]   (C)     at   (0,1);
 \draw[thick](A)--(B);
\draw[thick,dotted](B)--(C);
 \end{tikzpicture} 
  \ ,
 \end{align}
 and  we give the corresponding diagrammatic form  below each term. All  terms apart from the last arise either from the derivative, $\partial_C$,  hitting a $\cP$ (which we denote with a solid line) or commuting through (leaving an isolated vertex at $C$). The last term arises from the derivative, $\partial_C$, hitting the $\hat X.\partial$ term associated with the solid line between $A$ and $B$. We denote this by a dotted line from $C$ to $B$. Thus a solid line with a dotted line attached to the top of it loses its decoration, $\hat X.\partial$.

For the general case of several derivatives acting at a point we can work recursively: 
each additional  derivative either commutes through everything, corresponding to an isolated vertex, or it hits a $\cP$ corresponding to a solid line, or it hits a $\hat X.\partial$, denoted by a dotted line. We add all such lines in all possible ways. So the $n$-derivative term is given diagrammatically by summing all graphs containing $n$ vertices in a vertical line, with  any number of solid edges between any two points, such that no vertex is attached to more than one solid edge,  and with any number of dotted edges from the vertex at the top of a solid edge to a higher vertex either isolated or at the bottom of a solid edge.

 The above examples are already enough to illustrate the key ingredients of the general algorithm for obtaining an explicit expression for several covariant derivatives at a point, $\del_{A_1}\del_{A_2} \dots \del_{A_n}$ by summing over all possible graphs.
 \subsubsection*{Algorithm for $\del_{A_1}\del_{A_2} \dots \del_{A_n} \phi$}
   \begin{enumerate}
 	\item Draw $n$ vertices vertically. Each corresponds to an embedding space index ordered so  the bottom one corresponds to  $A_n$ and the top one to $A_1$.
 	\item Draw any number of solid edges between any two vertices such that each vertex is connected to at most one solid edge. 
 	\item Draw any number of dotted edges from the upper  vertex of a solid  edge up to 
 	either a higher  disconnected vertex or a higher vertex that is the lower vertex of a solid edge. No vertex can be attached to  more than one dotted edge.
 	\item Sum over all the resulting graphs with the following interpretation:
 	\begin{align}
 	\begin{tikzpicture}[scale=1.50]
 			\coordinate[dot,label=left:{$\scriptstyle A$}]   (A)     at   (0,0);
 		\end{tikzpicture}
 		\raisebox{0.15cm}{$ \hskip.15cm = \cP_A^{A'}\partial_{A'} \hskip 2.55cm $}
 	\raisebox{-0.15cm}{	\begin{tikzpicture}[scale=1.50]
 			\coordinate[dot,label=left:{$\scriptstyle A$}]   (A)     at   (0,0);
 			\coordinate[dot,label=left:{$\scriptstyle B$}]   (B)     at   (0,1/2);
 				\draw[thick](A)--(B);
 		\end{tikzpicture}
 	}
 		\raisebox{0.15cm}{$ \hskip.15cm = \cP_{AB} \hat X .\partial \hskip 2.55cm $}
 			\raisebox{-0.15cm}{	\begin{tikzpicture}[scale=1.50]
 				\coordinate[dot,label=left:{$\scriptstyle A$}]   (A)     at   (0,0);
 				\coordinate[dot,label=left:{$\scriptstyle B$}]   (B)     at   (0,1/2);
 				\coordinate   (C)     at   (0,1);
 				\draw[thick](A)--(B);
 				\draw[thick](A)--(B);
 				 \draw[thick,dotted](B)--(C);
 			\end{tikzpicture}
 		}
 		\raisebox{0.15cm}{$ \hskip.15cm = \cP_{AB} \hskip 2.55cm $}
 	\end{align}
 So solid edges come with a decoration $\hat X .\partial$ unless they have a dotted line attached to the top in which case the decoration is removed. (Otherwise the dotted lines can be ignored.)

  \end{enumerate}

 Now a derivative interaction term  consists of covariant derivatives acting on different scalars with indices contracted together pairwise. This we denote graphically by putting together two or more of the above vertical graphs and adding grey edges corresponding to the contractions. 
 So for example we obtain $\del_B\del_A\phi_1\, \del_B\del_A\phi_2$ by taking two copies of all the  two-derivative diagrams~\eqref{del2} and gluing the corresponding vertices together 
 \begin{align}\label{eg1}
 	\raisebox{0.6cm}{$\del_B\del_A\phi_1 \,\del_B\del_A\phi_2 \quad = \qquad $}&
 	\begin{tikzpicture}[scale=1.80]
 	\coordinate[dot,label=below:{$\scriptstyle \phi_1$}]   (A1)     at   (0,0);
 	\coordinate[dot]   (B1)     at   (0,1/2);
	\coordinate[dot,label=below:{$\scriptstyle \phi_2$}]   (A2)     at   (1/2,0);
	\coordinate[dot]   (B2)     at   (1/2,1/2);	
	\draw[very thin, gray](A1)--(A2);
	\draw[very thin, gray](B1)--(B2);
 	\end{tikzpicture} 
 	\raisebox{0.6cm}{$\quad + \quad  $}
 		\begin{tikzpicture}[scale=1.80]
 	\coordinate[dot,label=below:{$\scriptstyle \phi_1$}]   (A1)     at   (0,0);
 	\coordinate[dot]   (B1)     at   (0,1/2);
 	\coordinate[dot,label=below:{$\scriptstyle \phi_2$}]   (A2)     at   (1/2,0);
 	\coordinate[dot]   (B2)     at   (1/2,1/2);	
 	\draw[very thin, gray](A1)--(A2);
 	\draw[very thin, gray](B1)--(B2);
 	\draw[thick](A2)--(B2);
 	\end{tikzpicture} 
 		\raisebox{0.6cm}{$\quad + \quad  $}
 		\begin{tikzpicture}[scale=1.80]
 	\coordinate[dot,label=below:{$\scriptstyle \phi_1$}]   (A1)     at   (0,0);
 	\coordinate[dot]   (B1)     at   (0,1/2);
 	\coordinate[dot,label=below:{$\scriptstyle \phi_2$}]   (A2)     at   (1/2,0);
 	\coordinate[dot]   (B2)     at   (1/2,1/2);	
 	\draw[very thin, gray](A1)--(A2);
 	\draw[very thin, gray](B1)--(B2);
 	\draw[thick](A1)--(B1);
 	\end{tikzpicture} 
 		\raisebox{0.6cm}{$\quad + \quad  $}
 		\begin{tikzpicture}[scale=1.80]
 	\coordinate[dot,label=below:{$\scriptstyle \phi_1$}]   (A1)     at   (0,0);
 	\coordinate[dot]   (B1)     at   (0,1/2);
 	\coordinate[dot,label=below:{$\scriptstyle \phi_2$}]   (A2)     at   (1/2,0);
 	\coordinate[dot]   (B2)     at   (1/2,1/2);	
 	\draw[very thin, gray](A1)--(A2);
 	\draw[very thin, gray](B1)--(B2);
 	\draw[thick](A1)--(B1);
 	\draw[thick](A2)--(B2);
 	\end{tikzpicture} \notag\\
 =\qquad &  \cP^{AB}\cP^{CD} (\partial_A \partial_C \phi_1)  (\partial_B \partial_D \phi_2) +  \cP^{AB} (\partial_A \partial_B \phi_1) \hat X.\partial \phi_2 \notag \\ 
 	&+  \cP^{AB} (\partial_A \partial_B \phi_2) \hat X.\partial \phi_1 + \cP^A_A (\hat X.\partial \phi_1) (\hat X.\partial \phi_2) \ .
 \end{align}
 Similarly we obtain $\del_C\del_B\del_A\phi_1\, \del_A\phi_2\, \del_B\phi_3\, \del_C\phi_4$ by taking the three-derivative diagram~\eqref{del3} together with three more vertices to the right and gluing the vertices correspondingly 
 \begin{align}\label{4derivex}
 &{ \del^C \del^B \del^A \phi_1 \,\del_A\phi_2 \,\del_B\phi_3 \,\del_C\phi_4}\notag \\
\raisebox{0.9cm}{$\hskip.2cm= \hskip .2cm $}& \begin{tikzpicture}[scale=1.60]
 \coordinate[dot,label=below:{$\scriptstyle \phi_1$}]   (A)     at   (0,0);
 \coordinate[dot]   (B)     at   (0,1/2);
 \coordinate[dot]   (C)     at   (0,1);
 \coordinate[dot,label=below:{$\scriptstyle \phi_2$}]   (A2)     at   (1/2,0);
 \coordinate[dot,label=below:{$\scriptstyle \phi_3$}]   (B3)     at   (1,0);
 \coordinate[dot,label=below:{$\scriptstyle \phi_4$}]   (C4)     at   (3/2,0);
 \draw[very thin, gray](A)--(A2);
 \draw[very thin, gray](B)--(B3);
 \draw[very thin, gray](C)--(C4);
 \end{tikzpicture}
 \raisebox{0.9cm}{$\hskip-.2cm+ \hskip .2cm $}
 \begin{tikzpicture}[scale=1.60]
 \coordinate[dot,label=below:{$\scriptstyle \phi_1$}]   (A)     at   (0,0);
 \coordinate[dot]   (B)     at   (0,1/2);
 \coordinate[dot]   (C)     at   (0,1);
 \coordinate[dot,label=below:{$\scriptstyle \phi_2$}]   (A2)     at   (1/2,0);
 \coordinate[dot,label=below:{$\scriptstyle \phi_3$}]   (B3)     at   (1,0);
 \coordinate[dot,label=below:{$\scriptstyle \phi_4$}]   (C4)     at   (3/2,0);
 \draw[very thin, gray](A)--(A2);
 \draw[very thin, gray](B)--(B3);
 \draw[very thin, gray](C)--(C4);
 \draw[thick](A)..controls (-.15,1/2)..(C);
 \end{tikzpicture} 
 \raisebox{0.9cm}{$\hskip-.2cm+ \hskip .2cm $}
 \begin{tikzpicture}[scale=1.60]
 \coordinate[dot,label=below:{$\scriptstyle \phi_1$}]   (A)     at   (0,0);
 \coordinate[dot]   (B)     at   (0,1/2);
 \coordinate[dot]   (C)     at   (0,1);
 \coordinate[dot,label=below:{$\scriptstyle \phi_2$}]   (A2)     at   (1/2,0);
 \coordinate[dot,label=below:{$\scriptstyle \phi_3$}]   (B3)     at   (1,0);
 \coordinate[dot,label=below:{$\scriptstyle \phi_4$}]   (C4)     at   (3/2,0);
 \draw[very thin, gray](A)--(A2);
 \draw[very thin, gray](B)--(B3);
 \draw[very thin, gray](C)--(C4);
 \draw[thick](B)--(C);
 \end{tikzpicture} 
 \raisebox{0.9cm}{$\hskip-.2cm+ \hskip .2cm$}
 \begin{tikzpicture}[scale=1.60]
 \coordinate[dot,label=below:{$\scriptstyle \phi_1$}]   (A)     at   (0,0);
 \coordinate[dot]   (B)     at   (0,1/2);
 \coordinate[dot]   (C)     at   (0,1);
 \coordinate[dot,label=below:{$\scriptstyle \phi_2$}]   (A2)     at   (1/2,0);
 \coordinate[dot,label=below:{$\scriptstyle \phi_3$}]   (B3)     at   (1,0);
 \coordinate[dot,label=below:{$\scriptstyle \phi_4$}]   (C4)     at   (3/2,0);
 \draw[very thin, gray](A)--(A2);
 \draw[very thin, gray](B)--(B3);
 \draw[very thin, gray](C)--(C4);
 \draw[thick](A)--(B);
 \end{tikzpicture} 
 \raisebox{0.9cm}{$\hskip-.2cm+ \hskip .2cm $}
 \begin{tikzpicture}[scale=1.60]
 \coordinate[dot,label=below:{$\scriptstyle \phi_1$}]   (A)     at   (0,0);
 \coordinate[dot]   (B)     at   (0,1/2);
 \coordinate[dot]   (C)     at   (0,1);
 \coordinate[dot,label=below:{$\scriptstyle \phi_2$}]   (A2)     at   (1/2,0);
 \coordinate[dot,label=below:{$\scriptstyle \phi_3$}]   (B3)     at   (1,0);
 \coordinate[dot,label=below:{$\scriptstyle \phi_4$}]   (C4)     at   (3/2,0);
 \draw[very thin, gray](A)--(A2);
 \draw[very thin, gray](B)--(B3);
 \draw[very thin, gray](C)--(C4);
 \draw[thick](A)--(B);
 \draw[thick,dotted](B)--(C);
 \end{tikzpicture} \notag\\
=&  \cP^{AA'}\cP^{BB'}\cP^{CC'} (\partial_A \partial_B\partial_C \phi_1)  (\partial_{A'} \phi_2) (\partial_{B'} \phi_3)(\partial_{C'} \phi_4) +
\cP^{BB'}\cP^{AC} (\hat X .\partial \partial_B \phi_1)  (\partial_{A} \phi_2) (\partial_{B'} \phi_3)(\partial_{C} \phi_4) \notag \\
 &+
\cP^{AA'}\cP^{BC} (\hat X .\partial \partial_A \phi_1)  (\partial_{A'} \phi_2) (\partial_{B} \phi_3)(\partial_{C} \phi_4)+
\cP^{CC'}\cP^{AB} (\hat X .\partial \partial_C \phi_1)  (\partial_{A} \phi_2) (\partial_{B} \phi_3)(\partial_{C'} \phi_4) \notag \\
&+ 
\cP^{CC'}\cP^{AB} ( \partial_C \phi_1)  (\partial_{A} \phi_2) (\partial_{B} \phi_3)(\partial_{C'} \phi_4)
 \ .
 \end{align}
 
 The general algorithm for  interaction terms is then a straightforward extension of the one above for covariant derivatives acting on a single scalar.
 
 \subsubsection*{Algorithm for contact interactions in AdS}
 
 \begin{enumerate}
 	\item For each scalar $\phi_i$ with $n_i$ covariant derivatives acting on it, draw all the corresponding contributing vertical graphs using the above algorithm. Place the graphs for each scalar next to each other horizontally (taking the outer product over the  list of graphs at each point).
 	\item Draw grey lines between corresponding contracted vertices in the interaction term.
 	\item Finally sum over all the resulting graphs with the following interpretation:
 	\item Each  connected path of solid and grey lines with end points in the vertical line $\phi_i$ and $\phi_j$ corresponds to 
 	$\cP^{AB} \partial_A\phi_i \partial_B \phi_j$.
 	\item  Each solid line above $\phi_i$  corresponds to $\hat X.\partial \phi_i$, \em{as long as it doesn't have  a dotted line attached to its upper vertex}. (If it does have such a dotted line it has no additional contribution.) 
 \end{enumerate}
See the above two examples~\eqref{eg1} and~\eqref{4derivex}.
 
 So far we have only discussed AdS covariant derivatives. The above rules can be used with the obvious modifications if instead we are viewing the action on a sphere (i.e. $A,B$ indices become $I,J$ indices, $\hat X \rightarrow \hat Y$  and  $\cP^{AB} \rightarrow \cP^{IJ}$ in~\eqref{pdef}).  But our main purpose here is to consider AdS$\times$S covariant derivatives. Thus each vertex now represents a 10d index $\mu=(A,I)$, but there needs to be some non-trivial re-interpretation in the case of the product geometry.
 
 \subsubsection*{Algorithm for contact interactions in AdS$\times$S}
 
 The first three steps of the algorithm are as for the AdS case above. Then
 \begin{enumerate}
 	\setcounter{enumi}{3}
\item Each  connected path of solid and grey lines with end points in the vertical line above $\phi_i$ and $\phi_j$ respectively corresponds to 
 	$\cP^{\mu \nu} \partial_{\mu}\phi_i \partial_{\nu} \phi_j$, but: 
 \item 
 	Each solid line above $\phi_i$, {\em{as long as it doesn't have  a dotted line attached to its upper vertex}}, breaks this manifest 10d structure by   contributing a multiplicative factor  $\hat X^A\partial_A \phi_i$, 
 	if the index running through it is in AdS or $ - \hat Y^I\partial_I \phi_i$ if the index running through is in the sphere. (The minus sign appears in the latter case, since this term arises from a derivative hitting $\cP$ in~\eqref{pdef} which has a minus sign in for the internal case.)
 	
 	\item Finally there is an additional  subtlety related to the dotted lines. The dotted line ties together the index type corresponding to the otherwise potentially disconnected parts of the graph, and then contributes a factor of $+1$ if the index running through is in AdS or $-1$ if the index running through is in the sphere.
 	(Recall that the dotted lines arise from derivatives $\partial_{\hat X}$ or $\partial_{\hat Y}$ hitting the decoration $\hat X^A\partial_A \phi_i$ or $- \hat Y^I\partial_I \phi_i$. Thus firstly,  this  vanishes unless, the derivative type (AdS or S) is the same  as that of the solid line (hence tying together the index type), and secondly it gives $\pm 1$ depending on whether it is AdS or S.) 	
 \end{enumerate}
 
 Thus for example the AdS$\times$S  covariant version of~\eqref{4derivex} 
is, with each of the five lines corresponding to the five graphs in~\eqref{4derivex}
 \begin{align}
&  \del^\rho \del^\nu \del^\mu \phi_1 \,\del_\mu\phi_2 \,\del_\nu\phi_3 \,\del_\rho\phi_4 \notag \\
 =&  \cP^{\mu\mu'}\cP^{\nu\nu'}\cP^{\rho\rho'} (\partial_\mu \partial_\nu\partial_\rho \phi_1)  (\partial_{\mu'} \phi_2) (\partial_{\nu'} \phi_3)(\partial_{\rho'} \phi_4) \notag \\
 &+
 \cP^{\nu\nu'}\cP^{AC} (\hat X .\partial_{\hat X} \partial_\nu \phi_1)  (\partial_{A} \phi_2) (\partial_{\nu'} \phi_3)(\partial_{C} \phi_4)-\cP^{\nu\nu'}\cP^{IK} (\hat Y .\partial_{\hat Y} \partial_\nu \phi_1)  (\partial_{I} \phi_2) (\partial_{\nu'} \phi_3)(\partial_{K} \phi_4) \notag \\
 &+
 \cP^{\mu\mu'}\cP^{BC} (\hat X .\partial_{\hat X} \partial_\mu \phi_1)  (\partial_{\mu'} \phi_2) (\partial_{B} \phi_3)(\partial_{C} \phi_4)-\cP^{\mu\mu'}\cP^{JK} (\hat Y .\partial_{\hat Y} \partial_\mu \phi_1)  (\partial_{\mu'} \phi_2) (\partial_{J} \phi_3)(\partial_{K} \phi_4)\notag\\&+
 \cP^{\rho\rho'}\cP^{AB} (\hat X .\partial_{\hat X} \partial_\rho \phi_1)  (\partial_{A} \phi_2) (\partial_{B} \phi_3)(\partial_{\rho'} \phi_4) - \cP^{\rho\rho'}\cP^{IJ} (\hat Y .\partial_{\hat Y} \partial_\rho \phi_1)  (\partial_{I} \phi_2) (\partial_{J} \phi_3)(\partial_{\rho'} \phi_4) \notag \\
 &+ 
 \cP^{CC'}\cP^{AB} ( \partial_C \phi_1)  (\partial_{A} \phi_2) (\partial_{B} \phi_3)(\partial_{C'} \phi_4)-\cP^{KK'}\cP^{IJ} ( \partial_K \phi_1)  (\partial_{I} \phi_2) (\partial_{J} \phi_3)(\partial_{K'} \phi_4)
 \ .
\end{align}
In particular, note that only the first line is manifestly 10d covariant (has only 10d $\mu,\nu$ indices). Also compare carefully the penultimate with the  final line. These arise from similar graphs (the last two in~\eqref{4derivex} ) but one with a dotted line and one without.  In the final line, as well as the decoration $\hat X .\partial_{\hat X}$ or $\hat Y .\partial_{\hat Y}$ being absent,  the dotted line has tied together the two otherwise disconnected parts of the graph, meaning for example that all indices are either AdS or S, with no mixed ones, unlike the penultimate line.
 
 Finally, note that in practice for our purposes here, the derivatives will always be acting on bulk to  boundary propagators~\eqref{bbprop} and thus  partial derivatives acting on a single scalar $\del_{\mu_1}\del_{\mu_2}\dots \del_{\mu_{n_i}}\phi_i$,  gives $(-1)^{n_i} (\Delta_i)_{n_i} X^{\mu_1}..X^{\mu_{n_i}}$ etc.

\subsection{Mellin space}

The previous subsection gave an algorithm for obtaining explicit expressions for the integrands of generalised Witten diagrams in AdS$\times$S coming from contact interactions with derivatives.
This will result in integrands corresponding to decorations of the (no-derivative) contact diagram $D$~\eqref{adsxscont}. The decorations are in the form of polynomials in $X_{i}.X_j$, $Y_{i}.Y_j, Q_i$ and $P_i$  which are homogeneous at each point (i.e. scale the same  under the local scaling $X_{i}.X_j \rightarrow e_ie_j X_{i}.X_j $, $Y_{i}.Y_j\rightarrow e_ie_j Y_{i}.Y_j, Q_i\rightarrow e_i Q_i$ and $P_i\rightarrow e_i P_i$ ).
Each term of such a decoration thus has the form
\begin{align}\label{decoration}
 \frac1{4!}\frac{\prod_i \mathcal{C}_{\Delta_i}}{(-2)^{2\Sigma_\Delta}}\int_{\text{AdS}\times\text{S}} {d^{d+1} \hat X} {d^{d+1} \hat Y} \times  \frac{(X_i.X_j)^{n^X_{ij}}(Y_i.Y_j)^{n^Y_{ij}}Q_i^{n^Q_i}P_i^{n^P_i}\times (\Delta_1)_{n_1}(\Delta_2)_{n_2}(\Delta_3)_{n_3}(\Delta_4)_{n_4}}{(P_1+Q_1)^{\Delta_1+n_1}(P_2+Q_2)^{\Delta_2+n_2}(P_3+Q_3)^{\Delta_3+n_3}(P_4+Q_4)^{\Delta_4+n_4}} \ ,
\end{align}
with $n_i = n_i^P+n_i^Q+\sum_j n^X_{ij}+\sum_j n^Y_{ij}$. We define $\Sigma_X,\Sigma_Y$ to represent the sum of all the $n^X_{ij},n^Y_{ij}$ respectively, $\Sigma_Q, \Sigma_P$ represents {\em half} the sum of all the $n^Q_{i},n^P_{i}$ and $\Sigma_n$ half the sum of the $n_i$, so $\Sigma_n=\Sigma_P+\Sigma_Q+\Sigma_X+\Sigma_Y$. Such a decorated integral  will modify~\eqref{DinDB} to 
\begin{align}\label{MinDB}
(-2)^{2\Sigma_X+2\Sigma_Y}	\sum_{p_i=0}^\infty \prod_{i=1}^4(-1)^{p_i}\frac{(p_i+1)_{\Delta_i+n_i-1}}{\Gamma(\Delta_i)} (X_i.X_j)^{n^X_{ij}}(Y_i.Y_j)^{n^Y_{ij}} D^{(d)}_{p_i+\Delta_i+n_i-n^P_i}(X_i)B_{p_i+n^Q_i}(Y_i)\ .
\end{align}
Inserting the Mellin transform of $D$~\eqref{Ddef} and expansion of $B$~\eqref{bsum} and performing some re-definitions and simplifications then gives the Mellin amplitude (defined in~\eqref{mellinampdef}): 
\begin{align}\label{decorationmellin}
&\cM_{\Delta_i}[\eqref{decoration}] = 
	(-2)^{\Sigma_X}2^{\Sigma_Y}(-1)^{2\Sigma_Q}\times\notag\\
&\left(\prod_{i<j}{(\delta_{ij})_{n^X_{ij}}(d_{ij}{-}n^Y_{ij}{+}1)_{n^Y_{ij}}}{}\right)  
\left(\prod_i
{(p_i{+}n_i^X{+}\Delta_i)_{n^P_i}}{(p_i{-}n^Q_i{-}n^Y_i{+}1)_{n^Q_i}}
\right)
 {(\Sigma_p{-}\Sigma_Y{+}\tfrac d2{+}1)_{\Sigma_\Delta-d-1+\Sigma_X+\Sigma_Y}}\ ,\notag\\
 & \qquad \qquad \text{where} \qquad \sum_{i\neq j} \delta_{ij}=p_j+\Delta_j \qquad \qquad \qquad 	\sum_{i\neq j} d_{ij}=p_j\ .
\end{align}
We will use this general formula, in conjunction with the algorithm of the previous subsection to compute higher order terms in the $\alpha'$ expansion of $1/2$-BPS correlators in the next sections.

\section{$\alpha'^5$ corrections}\label{sec:alpha5}

After $\alpha'^3$, the next terms in the effective action for string corrections occur at $\alpha'^5$. In the flat space limit, such terms contain four derivatives, so first we wish to consider all the possible terms in the effective action on AdS$\times$S involving four derivatives.
At first there are many terms one can write down, but then using integration by parts as well as the equations of motion reduces the number down quickly. We find that in fact there are only two linearly independent terms one can write down involving four derivatives:
\begin{align}\label{inda5}
\left(\nabla\phi.\nabla\phi\right)^{2} \qquad \text{and} \qquad \nabla^{2}\nabla_{\mu}\phi\nabla^{\mu}\phi\phi^{2}\ .
\end{align}
These are the two terms appearing in the effective action~\eqref{Seff}. Any  other four-derivative term can be written in terms of these, using integration by parts and the equations of motion.  For example
\begin{align}
&\nabla_{\mu}\nabla_{\nu}\phi\nabla^{\mu}\phi\nabla^{\nu}\phi\sim-\frac{1}{2}\left(\nabla\phi.\nabla \phi\right)^{2} \ ,\notag\\
&\left(\nabla_\mu\nabla_\nu\phi\nabla^\mu\nabla^\nu\phi\right)\phi^{2}\sim\left(\nabla\phi.\nabla\phi\right)^{2}-\nabla^{2}\nabla_{\mu}\phi\nabla^{\mu}\phi\phi^{2}\ .
\end{align}
Although at this level the independent integrands can be obtained by hand,  they can also be nicely checked on a computer by using the algorithm of the previous section and converting to Mellin space where the IBP identities are  made manifest. Simply list all possible four-derivative integrands on the computer, use the algorithm to obtain the corresponding integrand, convert them to Mellin amplitudes,  and then solve for the independent ones.

We see here for the first time that the effective action has an ambiguity - a term not determined by the Virasoro-Shapiro amplitude: in the flat space limit the second integrand in~\eqref{inda5} will vanish (as we can commute the Laplacian through so it acts directly on $\phi$ giving zero by the equations of motion) and so remains undetermined. The complete effective action at this order is thus (see~\eqref{Seff})
\begin{align}
S_{\alpha'^5}=\frac18\left( \frac{\alpha'}2\right)^5 \left(\zeta_5 S_{\alpha'^5}^{\text{main}} + C_0 S_{\alpha'^5}^{\text{amb}}+ A_2 S_{\alpha'^3}^{\text{main}}\right)\ ,
\label{alphap5corr}
\end{align}
where
\begin{align}\label{4deriv}
S_{\alpha'^5}^{\text{main}} &=  \frac3{4!}\int_{\text{AdS}\times\text{S}} {d^{5} {\hat X}} {d^{5} {\hat Y}}
(\nabla\phi.\nabla\phi)(\nabla\phi. \nabla\phi)\ ,\notag\\
S_{\alpha'^5}^{\text{amb}} &=  \frac6{4!}\int_{\text{AdS}\times \text{S}} d^5\hat X d^5 \hat Y    \nabla^{2}\nabla_{\mu}\phi\nabla^{\mu}\phi\phi^{2}\ ,\notag\\
S_{\alpha'^3}^{\text{main}} &=  \frac1{4!}\int_{\text{AdS}\times \text{S}} d^5\hat X d^5 \hat Y    \phi^4\ .
\end{align}

Replacing the scalar fields by bulk-to-boundary propagators and applying the covariant derivatives directly on them then gives a prediction for the $1/2$-BPS correlators at this order in $\alpha'$. First consider the main contribution~\eqref{4deriv}:
\begin{align}\label{ap5}
&\langle \cO\cO\cO\cO\rangle|_{\alpha'^5;\text{main}}&\notag \\ &=\frac1{4!}\frac {(\mathcal{C}_4)^4}{(-2)^{16}} \int_{\text{AdS}\times \text{S}} {d^{5} {\hat X}} {d^{5} {\hat Y}} \frac{N_{12}N_{34}+N_{13}N_{24}+N_{14}N_{23}}{(P_1+Q_1)^{5}(P_2+Q_2)^{5}(P_3+Q_3)^{5}(P_4+Q_4)^{5}}\times 4^4 \ ,
 \end{align}  
where 
\begin{equation}\label{Nij}
N_{ij} = X_i.X_j+Y_i.Y_j +P_iP_j-Q_iQ_j \ .
\end{equation}
This can then be straightforwardly expanded to give any correlator directly and explicitly in position space in terms of AdS and S contact diagram functions,  as is done for a general integral in~\eqref{MinDB}.
The corresponding Mellin amplitude can also  be read off directly from~\eqref{decorationmellin}
\begin{align}\label{4dervmellinFA}
\mathcal{M}_{\alpha'^5}^{\rm{main}}=4\,\Big[&\left(\Sigma_p{{-}}1\right)_5\left({\mathbf s} ^2+{\mathbf t}^2+{\mathbf u}^2\right)\nonumber\\
+&\left(\Sigma_p{{-}}1\right)_4 \left(-10\left(\tilde{s}\,{\mathbf s}+\tilde{t}\,{\mathbf t}+\tilde{u}\,{\mathbf u}\right)-5\left(c_s\,{\mathbf s}+c_t\, {\mathbf t}+c_u\, {\mathbf u}\right)\right)\nonumber\\
+&\left(\Sigma_p{{-}}1\right)_3\left(20\,\left(\tilde{s}^2+\tilde{t}^2+\tilde{u}^2\right)+4\,\left(c_s^2+c_t^2+c_u^2\right)+20\left(\tilde{s}\,c_s+\tilde{t}\,c_t+\tilde{u}\,c_u\right)\right)\nonumber\\
+&\left(\Sigma_p{{-}}1\right)_3\left(-12\,\Sigma_p^2\right)\Big]\ .
\end{align}
Here we have used~\eqref{decorationmellin} to obtain the Mellin amplitude (with $\Delta_i=4,d=4$ and $p_i\rightarrow p_i-2$)  and then solved the constraints
$$\sum_i\delta_{ij} = p_j+2 \qquad \qquad \sum_id_{ij} = p_j-2$$
in terms of new variables $(s,t,u)$ and $(\tilde s, \tilde t, \tilde u)$, which are defined as  follows~\cite{Aprile:2020luw}:
\begin{align}
&\delta_{12}=-s+c_s \ ,\quad &&\delta_{14}=-t+c_t \ ,\quad &&\delta_{13}=-u \ ,\nonumber \\
&\delta_{23}=-t \ ,\quad &&\delta_{24}=-u+c_u \ ,\quad &&\delta_{34}=-s \ ,\nonumber \\
&d_{12}=\tilde{s}+c_s \ ,\quad &&d_{14}=\tilde{t}+c_t \ ,\quad &&d_{13}=\tilde{u} \ ,\nonumber\\
&d_{23}=\tilde{t} \ ,\quad &&d_{24}=\tilde{u}+c_u \ ,\quad &&d_{34}=\tilde{s} \ , \nonumber \\
&{\mathbf s}=s+\tilde{s} \ ,\quad &&{\mathbf t}=t+\tilde{t} \ ,\quad &&{\mathbf u}=u+\tilde{u} \ ,
\end{align}
where $s+t+u=-p_3-2\, ,\ \tilde{s}+\tilde{t}+\tilde{u}=p_3-2$ and ${\mathbf s}+{\mathbf t}+{\mathbf u}=-4$. We also define
\begin{equation}
c_s=\frac{p_1+p_2-p_3-p_4}{2} \ ,\quad c_t=\frac{p_1+p_4-p_2-p_3}{2} \ ,\quad c_u=\frac{p_2+p_4-p_3-p_1}{2} \ .
\end{equation}

Now let us take a closer look at the ambiguity in the second line of~\eqref{4deriv}.
Using the equations of motion in~\eqref{eom}, the integrand can be written
as
\begin{equation}
\nabla^{2}\nabla_{\mu}\phi\nabla^{\mu}\phi\phi^{2}=\left[\nabla_{\hat{X}}^{2},\nabla_{A}\right]\phi\nabla^{A}\phi+\left[\nabla_{\hat{Y}}^{2},\nabla_{I}\right]\phi\nabla^{I}\phi \ .
\end{equation}
Moreover, after some  algebra we find that
\begin{equation}
\left[\nabla_{\hat{X}}^{2},\nabla_{A}\right]\phi=-d\nabla_{A}\phi,\,\,\,\  \ \left[\nabla_{\hat{Y}}^{2},\nabla_{I}\right]\phi=d\nabla_{I}\phi \ ,
\end{equation}
so the ambiguity can be written as
\begin{equation}
\nabla^{2}\nabla_{\mu}\phi\nabla^{\mu}\phi\phi^{2}=-d\left(\left(\nabla_{\hat{X}}\phi\right)^{2}-\left(\nabla_{\hat{Y}}\phi\right)^{2}\right)\phi^{2}\ .
\end{equation}
The corresponding Witten diagram expression is given by 
\begin{equation}
\langle \cO\cO\cO\cO\rangle|_{\alpha'^5;\text{amb}}=-\frac1{4!}\frac {(\mathcal{C}_4)^4}{(-2)^{16}}\int_{\text{AdS}\times\text{S}}\frac{d^{5}\hat{X}d^{5}\hat{Y}}{\prod_{i}\left(P_{i}+Q_{i}\right)^{4}}\sum_{i<j}\frac{L_{ij}}{\left(P_{i}+Q_{i}\right)\left(P_{j}+Q_{j}\right)}\times 4^3 \ ,
\label{ap5amb}
\end{equation}
where
\begin{equation}
L_{ij}=X_{i}. X_{j}+P_{i}P_{j}-Y_{i}. Y_{j}+Q_{i}Q_{j}\ .
\end{equation}
This takes a very simple form in Mellin space 
\begin{equation}
\mathcal{M}_{\alpha'^5}^{\rm{amb}}=4\,(\Sigma_p-1)_3\left(c_s^2+c_t^2+c_u^2+\Sigma_p^2-16\right)\ .
\label{melap5amb}
\end{equation}

Moreover, after multiplying the $\alpha'^3$ term in~\eqref{melap3} by $(\alpha'/(2R^2))^2$ (where we set $R=1$), it can be thought of as an additional ambiguity at $\alpha'^5$, which is the origin of the third line in~\eqref{4deriv}. Restoring the prefactors in~\eqref{alphap5corr}, the $\alpha'^5$ correction to the Mellin amplitude for $1/2$-BPS correlators can be written as a sum over three terms: 
\begin{equation}\label{ap5span}
\cM_{\alpha'^5}=\frac18\left(\frac{\alpha'}{2}\right)^5\left(\zeta_5\mathcal{M}_{\alpha'^{5}}^{{\rm {main}}}+C_0\mathcal{M}_{\alpha'^{5}}^{{\rm {amb}}}+A_2\mathcal{M}_{\alpha'^{3}}^\text{main}\right)\ ,
\end{equation}
where $\mathcal{M}_{\alpha'^{3}}^\text{main}=(\Sigma_p-1)_3$ (it is given in~\eqref{melap3} but without the explicit normalisation there).
The coefficients of the subleading terms can be fixed by comparing to the localisation result in~\cite{Binder:2019jwn} and are given by 
\begin{equation}
C_0=-\frac{3}{2}\,\zeta_5\ ,\quad A_2=-30\,\zeta_5\ .
\label{alpha5coeffs}
\end{equation}
We find perfect agreement with the results from bootstrap methods of~\cite{Drummond:2020dwr} (rewritten in this notation in~\cite{Aprile:2020luw})%
\footnote{We thank Francesco Aprile for explicitly checking this agreement.}.

\section{$\alpha'^6$ corrections}\label{sec:alpha6}

At order $\alpha'^6$ we have to consider all possible terms in the effective action involving six derivatives. Using a computer, it is straightforward to enumerate all possibilities and compute their Mellin amplitudes using the algorithm explained in section~\ref{wdiagal} to find all linearly independent terms. After doing so, we find that there are only two linearly independent terms involving six derivatives:
\begin{equation}\label{inda6}
    (\nabla\phi.\nabla\phi)(\nabla_\mu\nabla_\nu \phi \nabla^\mu\nabla^\nu\phi)\quad \text{and}\quad \nabla_{\mu}\nabla^{2}\nabla_{\nu}\phi\nabla^{\mu}\nabla^{\nu}\phi\phi^{2} \ ,
\end{equation}
which appear in the effective action in~\eqref{Seff}. The first term is the main correction at $\alpha'^6$ while the second term is an ambiguity which vanishes in the flat space limit and is thus not determined by the flat space Virasoro-Shapiro amplitude.

The complete action at order $\alpha'^6$ is given by (see~\eqref{Seff})
\begin{align}
S_{\alpha'^6}=\frac18\left( \frac{\alpha'}2\right)^6 \left(2(\zeta_3)^2 S_{\alpha'^6}^{\text{main}} +E_0 S_{\alpha'^6}^\text{amb}+B_1 S_{\alpha'^5}^\text{main}+ C_1 S_{\alpha'^5}^{\text{amb}}+ A_3 S_{\alpha'^3}^{\text{main}}\right)\ ,
\end{align}
where
\begin{align}\label{6deriv}
S_{\alpha'^6}^{\text{main}} &=  \frac1{4!}\int_{\text{AdS}\times \text{S}} {d^{5} {\hat X}} {d^{5} {\hat Y}}
(\nabla\phi.\nabla\phi)(\nabla_\mu\nabla_\nu \phi \nabla^\mu\nabla^\nu\phi)\ ,\notag\\
S_{\alpha'^6}^{\text{amb}} &=  \frac6{4!}\int_{\text{AdS}\times \text{S}} d^5\hat X d^5 \hat Y    \nabla_{\mu}\nabla^{2}\nabla_{\nu}\phi\nabla^{\mu}\nabla^{\nu}\phi\phi^{2} \ ,
\end{align}
and the rest was defined in~\eqref{4deriv} (in particular, they arise from taking all the terms contributing at $\alpha'^5$ and multiplying them with $\alpha'/(2R^2)$ with unfixed numerical coefficients (note that we set $R=1$)). We then find that the main contribution to $1/2$-BPS correlators at this order is
\begin{align}\label{ap6}
&\langle \cO\cO\cO\cO\rangle|_{\alpha'^6;\text{main}}&\notag \\ &=\frac1{4!}\frac{1}{6}\frac {(\mathcal{C}_4)^4}{(-2)^{16}} \int_{\text{AdS}\times \text{S}} \frac{d^{5}\hat{X}d^{5}\hat{Y}}{\prod_{i}\left(P_{i}+Q_{i}\right)^{5}}\left[\frac{N_{12}M_{34}}{\left(P_{3}+Q_{3}\right)\left(P_{4}+Q_{4}\right)}+{\rm perms}\right]\times 4^4\times 5^2\ ,
 \end{align}  
where the correlator is understood to come from the first line of~\eqref{6deriv}, $N_{ij}$ was defined in~\eqref{Nij} and
\begin{equation}\label{Mij}
M_{ij}=\left(X_{i}. X_{j}+P_{i}P_{j}+Y_{i}. Y_{j}-Q_{i}Q_{j}\right)^{2}-\frac{1}{5}\left(P_{i}P_{j}-Q_{i}Q_{j}\right)^{2}\ .
\end{equation}

Before discussing the Mellin amplitude of the main contribution, let us briefly describe the ambiguity whose integrand can be written as
\begin{equation}
\nabla_{\mu}\nabla^{2}\nabla_{\nu}\phi\nabla^{\mu}\nabla^{\nu}\phi\phi^{2}=-d\left(\left(\nabla_{A}\nabla_{B}\phi\right)^{2}-\left(\nabla_{I}\nabla_{J}\phi\right)^{2}\right)\phi^{2}\ ,
\label{6dervamb}
\end{equation}
where $A$ and $I$ indices label $\hat{X}$ and $\hat{Y}$ coordinates, respectively. We obtained the right hand side by commuting the $\nabla^2$ with $\nabla_\nu$ and using the equations of motion as we did in the previous subsection. The Witten diagram expression associated with~\eqref{6dervamb} is
\begin{align}\label{ap6amb}
&\langle \cO\cO\cO\cO\rangle|_{\alpha'^6;\text{amb}}&\notag \\ &=\frac1{4!}\frac {(\mathcal{C}_4)^4}{(-2)^{16}} \int_{\text{AdS}\times \text{S}} \frac{d^{5}\hat{X}d^{5}\hat{Y}}{\prod_{i}\left(P_{i}+Q_{i}\right)^{4}}\sum_{i<j}\frac{K_{ij}}{\left(P_{i}+Q_{i}\right)^{2}\left(P_{j}+Q_{j}\right)^{2}}\times 4^3\times 5^2\ ,
\end{align}
where
\begin{equation}
K_{ij}=\left(X_{i}. X_{j}+P_{i}P_{j}\right)^{2}-\left(Y_{i}. Y_{j}-Q_{i}Q_{j}\right)^{2}-\frac{1}{5}\left(\left(P_{i}P_{j}\right)^{2}-\left(Q_{i}Q_{j}\right)^{2}\right)\ .
\end{equation}
Converting this to Mellin space gives the ambiguity
\begin{align}\label{6dervmellinambiguity}
\mathcal{M}_{\alpha'^{6}}^{{\rm {amb}}}=&&-32\,\Big[\left(\Sigma_p{{-}}1\right)_5&\,\left({\mathbf s}^2+{\mathbf t}^2+{\mathbf u}^2\right)\nonumber\\
&&+\left(\Sigma_p{{-}}1\right)_4\,&\left(\frac{1}{2}\left({\mathbf s}\,c_s^2+{\mathbf t}\,c_t^2+{\mathbf u}\,c_u^2\right)-\left(\Sigma_p+3\right)\left[2\left({\mathbf s}\,\tilde{s}+{\mathbf t}\,\tilde{t}+{\mathbf u}\,\tilde{u}\right)+\left({\mathbf s}\,c_s+{\mathbf t}\,c_t+{\mathbf u}\,c_u\right)\right]\right)\nonumber \\    
&&+\left(\Sigma_p{{-}}1\right)_3\,&\Big(-\left(c_s^3+c_t^3+c_u^3\right)-2\left(c_s^2\,\tilde{s}+c_t^2\,\tilde{t}+c_u^2\,\tilde{u}\right)+10\,\Sigma_p\left(\tilde{s}^2+\tilde{t}^2+\tilde{u}^2\right)\nonumber\\
&&&\ \,+10\,\Sigma_p\left(\tilde{s}\,c_s+\tilde{t}\,c_t+\tilde{u}\,c_u\right)
+3\,\Sigma_p\left(c_s^2+c_t^2+c_u^2\right)-2\,\Sigma_p^3-16\,\Sigma_p\Big)\Big]\ .
\end{align}

Converting~\eqref{ap6} to Mellin space, the Mellin amplitude of the main contribution is
\begin{equation}
\mathcal{M}_{\alpha'^{6}}^\text{main}=  \hat{\mathcal{M}}_{\alpha'^{6}}^\text{main}-\frac1{12} \mathcal{M}_{\alpha'^6}^\text{amb},
\end{equation}
where 
\begin{align}\label{6dervmellinFA}
\hat{\mathcal{M}}_{\alpha'^{6}}^{{\rm {main}}}=&&\frac{8}{3}\,\Big[ \left(\Sigma_p{{-}}1\right)_6&\,\left({\mathbf s} ^3+{\mathbf t}^3+{\mathbf u}^3\right)\nonumber\\
&&+\left(\Sigma_p{{-}}1\right)_5\,&\left(6\,\Sigma_p\left({\mathbf s}^2+{\mathbf t}^2+{\mathbf u}^2\right)-18\left({\mathbf s}^2\,\tilde{s}+{\mathbf t}^2\,\tilde{t}+ {\mathbf u}^2\,\tilde{u}\right)-9\left({\mathbf s}^2\,c_s+{\mathbf t}^2\,c_t+{\mathbf u}^2\,c_u\right)\right)\nonumber\\
&&+\left(\Sigma_p{{-}}1\right)_4\,&\Big(90\left({\mathbf s}\,\tilde{s}^2+{\mathbf t}\,\tilde{t}^2+{\mathbf u}\,\tilde{u}^2\right)+\frac{39}{2}\left({\mathbf s}\,c_s^2+{\mathbf t}\,c_t^2+{\mathbf u}\,c_u^2\right)+90\left({\mathbf s}\,\tilde{s}\,c_s+{\mathbf t}\,\tilde{t}\,c_t+{\mathbf u}\,\tilde{u}\,c_u\right)\nonumber\\
&&&\ \,-60\,\Sigma_p\left({\mathbf s}\,\tilde{s}+{\mathbf t}\,\tilde{t}+{\mathbf u}\,\tilde{u}\right)-30\,\Sigma_p\left({\mathbf s}\,c_s+{\mathbf t}\,c_t+{\mathbf u}\,c_u\right)\Big)\nonumber \\ 
&&+\left(\Sigma_p{{-}}1\right)_3\,&\Big(-120\left(\tilde{s}^3+\tilde{t}^3+\tilde{u}^3\right)-9\left(c_s^3+c_t^3+c_u^3\right)-180\left(\tilde{s}^2\,c_s+\tilde{t}^2\,c_t+\tilde{u}^2\,c_u\right)\nonumber\\
&&&\ \,-78\left(c_s^2\,\tilde{s}+c_t^2\,\tilde{t}+c_u^2\,\tilde{u}\right)+120\,\Sigma_p\left(\tilde{s}^2+\tilde{t}^2+\tilde{u}^2\right)+27\,\Sigma_p\left(c_s^2+c_t^2+c_u^2\right)\notag\\
&&&\ \,+120\,\Sigma_p\left(\tilde{s}\,c_s+\tilde{t}\,c_t+\tilde{u}\,c_u\right)-50\,\Sigma_p^3-16\,\Sigma_p\Big)\Big]\ .
\end{align}
This Mellin amplitude shows a similar structure as~\eqref{4dervmellinFA}. Every line is multiplied by a Pochhammer depending on the power of $\{{\mathbf s},{\mathbf t},{\mathbf u}\}$ and the rest is at most cubic in the variables $\{{\mathbf s},{\mathbf t},{\mathbf u},\tilde{s},\tilde{t},\tilde{u},c_s,c_t,c_u,\Sigma_p\}$.

Additionally, after multiplying the three terms which span the $\alpha'^5$ correction in~\eqref{ap5span} by $\alpha'$, they become additional ambiguities at $\alpha'^6$, see the expansion~\eqref{coeffexpansion}.
The complete Mellin amplitude for $1/2$-BPS correlators at order $\alpha'^6$ can be written as a sum over five terms:
\begin{equation}\label{ap6span}
\cM_{\alpha'^6}=\frac18\left( \frac{\alpha'}2\right)^6 (2\,(\zeta_3)^2  \mathcal{M}_{\alpha'^{6}}^{{\rm {main}}}+E_0\mathcal{M}_{\alpha'^{6}}^{{\rm {amb}}}+B_1\mathcal{M}_{\alpha'^{5}}^{{\rm {main}}}+C_1\mathcal{M}_{\alpha'^{5}}^{{\rm {amb}}}+A_3\mathcal{M}_{\alpha'^{3}}^\text{main})\ ,
\end{equation}
where we restore the coefficients from~\eqref{ap6}.%
\footnote{Note that the number of ambiguities is consistent with the number  obtained via the  bootstrap method. We thank Francesco Aprile, James Drummond, Hynek Paul and Michele Santagata for discussions on this.} 

We can fix two of the coefficients by comparing the Mellin amplitude to the result from localisation in~\cite{Chester:2020dja,Chester:2020vyz}.
To compare~\eqref{ap6span} to~\cite{Chester:2020dja} we take $s\rightarrow\tfrac{s}{2}-2$, $t\rightarrow\tfrac{t}{2}-2$ and specialise to $p_i=2$ (where $\tilde{s}=\tilde{t}=\tilde{u}=0$):
\begin{align}\label{alpha62222}
\cM_{\alpha'^6}^{2222}=\frac18\left( \frac{\alpha'}2\right)^6\times60\,\Big(&672 (\zeta_3)^2\, s\,t\,u+14 \left(3\,B_1+4\left((\zeta_3)^2-6\,E_0\right)\right) \left(s^2 + t^2 + u^2\right)\notag\\
&+A_3-96\,B_1+768\,E_0-3200(\zeta_3)^2\Big)\ ,
\end{align}
where $u=4-s-t$.
We can now compare this expression to the result in~\cite{Chester:2020dja} and partially fix the coefficients to
\begin{equation}
E_0=\frac{B_1}{8}+\frac{2(\zeta_3)^2}{3}\ , \quad A_3=0\ ,
\label{alpha6coeffs}
\end{equation}
which leads to the $\alpha'^6$ correction to the correlator for $p_i=2$:
\begin{align}\label{alpha62222}
\cM_{\alpha'^6}^{2222}=\frac18\left( \frac{\alpha'}2\right)^6\times 2\,(\zeta_3)^2\times(3)_6\left[s\,t\,u-\frac{1}{4}  \left(s^2 + t^2 + u^2\right)-4\right]\ .
\end{align}
It is  noteworthy that localisation predicts the coefficient $A_3=0$. Localisation also predicts the absence of any $\alpha'^4$ corrections i.e. $A_1=0$~\cite{Binder:2019jwn}. Indeed, as we discuss in the conclusions, it is natural to expect all odd terms in the expansion of the coefficients in $\alpha'/R^2$ (see~\eqref{coeffexpansion}) to vanish in which case we would have  $B_1=C_1=0$ also, then the $\alpha'^6$ correction to the Mellin amplitude in \eqref{ap6span} is completely fixed.

\section{$\alpha'^7$ corrections}\label{sec:alpha7}

Using the algorithm explained in section~\ref{wdiagal} to find all linearly independent terms in the effective action involving eight derivatives, we find that there are six independent terms, notably the main contribution
\begin{equation}
    (\nabla_\mu\nabla_\nu \phi \nabla^\mu\nabla^\nu\phi)^2,
\end{equation}
and five ambiguities:
\begin{align}\label{alpha7ambs}
&\left(\nabla^\mu\nabla^\nu\nabla_\mu\nabla^\rho\nabla^\sigma\nabla_\rho\phi\right)\left(\nabla_\nu\nabla_\sigma\phi\right)\phi^2\ ,\quad  
\left(\nabla^2\nabla^\mu\nabla^\nu\nabla^\rho\nabla_\nu\phi\right)\left(\nabla_\mu\nabla_\rho\phi\right)\phi^2\ ,\notag\\
&\left(\nabla^2\nabla^\mu\nabla^\nu\nabla^\rho\nabla_\nu\phi\right)\left(\nabla_\rho\phi\right)\left(\nabla_\mu\phi\right)\phi \ ,\quad  \left(\nabla^\mu\nabla^\nu\nabla^\rho\nabla_\nu\nabla_\rho\phi\right)\left(\nabla^\sigma\nabla_\mu\phi\right)\left(\nabla_\sigma\phi\right)\phi\ ,\notag \\ &\left(\nabla^\mu\nabla^\nu\nabla^\rho\nabla^\sigma\nabla_\rho\phi\right)\left(\nabla_\mu\nabla_\sigma\phi\right)\left(\nabla_\nu\phi\right)\phi\ .
\end{align}
See appendix~\ref{appendix:alpha7ambiguities} for details on the ambiguities. The complete effective action at this order is then given by (see~\eqref{Seff})
\begin{align}\label{ap7}
S_{\alpha'^7}=\frac18\left(\frac{\alpha'}2\right)^7 \Big(&\tfrac{1}{2}\zeta_7 S_{\alpha'^7}^{\text{main}}+G_{1;0} S_{\alpha'^7}^{\text{amb}_1}+G_{2;0} S_{\alpha'^7}^{\text{amb}_2}+G_{3;0} S_{\alpha'^7}^{\text{amb}_3}+G_{4;0} S_{\alpha'^7}^{\text{amb}_4}+G_{5;0} S_{\alpha'^7}^{\text{amb}_5}\notag\\
&+D_1 S_{\alpha'^6}^\text{main}+E_1 S_{\alpha'^6}^\text{amb}+B_2 S_{\alpha'^5}^\text{main}+ C_2 S_{\alpha'^5}^{\text{amb}}+ A_4 S_{\alpha'^3}^{\text{main}}\Big) \ ,
\end{align}
where the main contribution is
\begin{align}\label{8deriv}
S_{\alpha'^7}^{\text{main}} =  \frac6{4!}\int_{\text{AdS}\times \text{S}} {d^{5} {\hat X}} {d^{5} {\hat Y}}
(\nabla_\mu\nabla_\nu \phi \nabla^\mu\nabla^\nu\phi)^2,
\end{align}
and the contributions from the five $\alpha'^7$ ambiguities in \eqref{alpha7ambs} to the effective action are given in appendix~\ref{appendix:alpha7ambiguities} together with their Witten diagram expressions and Mellin amplitudes. The contributions from lower $\alpha'$ orders were defined in~\eqref{4deriv} and~\eqref{6deriv}. The prediction for the main contribution to the $1/2$-BPS correlator at this order is:
\begin{align}\label{ap8}
\langle \cO\cO\cO\cO\rangle|_{\alpha'^7;\text{main}}=\frac2{4!}\frac {(\mathcal{C}_4)^4}{(-2)^{16}} \int_{\text{AdS}\times \text{S}} \frac{d^{5}\hat{X}d^{5}\hat{Y}}{\prod_i\left(P_{i}+Q_{i}\right)^{6}}\left[M_{12}M_{34}+{\rm perms}\right]\times 4^4\times 5^4\, ,
 \end{align} 
where $M_{ij}$ was defined in~\eqref{Mij}. The Mellin amplitude of the main contribution is
\begin{align}
\cM_{\alpha'^7}^\text{main}=&\,\hat{\cM}_{\alpha'^7}^\text{main}+\frac{63}{8} \cM_{\alpha'^7}^{\text{amb}_1}-\frac{31}{4} \cM_{\alpha'^7}^{\text{amb}_2}-\frac{25}{32} \cM_{\alpha'^7}^{\text{amb}_3}- \cM_{\alpha'^7}^{\text{amb}_4}-32 \cM_{\alpha'^5}^{\text{main}}\notag\\
&-\frac{85}{2} \cM_{\alpha'^5}^{\text{amb}}-1024 \cM_{\alpha'^3}^{\text{main}}\ ,
\end{align}
where $\hat{\cM}_{\alpha'^7}^\text{main}$ is:
{\small
\begin{align}
\hat{\mathcal{M}}_{\alpha'^{7}}^{{\rm {main}}}=&&32\,\Big[ \left(\Sigma_p{{-}}1\right)_7\,&\left({\mathbf s}^4+{\mathbf t}^4+{\mathbf u}^4\right) \notag\\
&&+\left(\Sigma_p{{-}}1\right)_6\,& \left(8\,\Sigma_p\left({\mathbf s}^3+{\mathbf t}^3+{\mathbf u}^3\right)-28\left({\mathbf s}^3\, \tilde{s}+{\mathbf t}^3\,\tilde{t}+{\mathbf u}^3\,\tilde{u}\right) -14\left({\mathbf s}^3\, c_s+{\mathbf t}^3\,c_t+{\mathbf u}^3\,c_u\right)\right) \notag\\
&&+\left(\Sigma_p{{-}}1\right)_5\,& \Big(\Sigma_p\left(26\,\Sigma_p+9\right)\left({\mathbf s}^2+{\mathbf t}^2+{\mathbf u}^2\right)+252\left({\mathbf s}^2\, \tilde{s}^2+{\mathbf t}^2\,\tilde{t}^2+{\mathbf u}^2\,\tilde{u}^2\right)\notag\\
&&&\ \,+252\left({\mathbf s}^2\, \tilde{s}\,c_s+{\mathbf t}^2\,\tilde{t}\,c_t+{\mathbf u}^2\,\tilde{u}\,c_u\right)+57\left({\mathbf s}^2\, c_s^2+{\mathbf t}^2\,c_t^2+{\mathbf u}^2\,c_u^2\right)\notag\\
&&&\ \,-144\,\Sigma_p\left({\mathbf s}^2\, \tilde{s}+{\mathbf t}^2\,\tilde{t}+{\mathbf u}^2\,\tilde{u}\right)-72\,\Sigma_p\left({\mathbf s}^2\, c_s+{\mathbf t}^2\,c_t+{\mathbf u}^2\,c_u\right)\Big)\notag\\
&&+\left(\Sigma_p{{-}}1\right)_4\,& \Big(-840\left({\mathbf s}\,\tilde{s}^3+{\mathbf t}\,\tilde{t}^3+{\mathbf u}\,\tilde{u}^3\right)-75\left({\mathbf s}\,c_s^3+{\mathbf t}\,c_t^3+{\mathbf u}\,c_u^3\right)\notag\\
&&&\ \,-1260\left({\mathbf s}\,\tilde{s}^2\,c_s+{\mathbf t}\,\tilde{t}^2\,c_t+{\mathbf u}\,\tilde{u}^2\,c_u\right) -570\left({\mathbf s}\,c_s^2\,\tilde{s}+{\mathbf t}\,c_t^2\,\tilde{t}+{\mathbf u}\,c_u^2\,\tilde{u}\right)\notag\\
&&&\ \,+720\,\Sigma_p\left[\left({\mathbf s}\,\tilde{s}^2+{\mathbf t}\,\tilde{t}^2+{\mathbf u}\,\tilde{u}^2\right) +\left({\mathbf s}\,\tilde{s}\,c_s+{\mathbf t}\,\tilde{t}\,c_t+{\mathbf u}\,\tilde{u}\,c_u\right)\right]\notag\\
&&&\ \,+\frac{1}{2}\left(336\,\Sigma_p-1\right)\left({\mathbf s}\,c_s^2+{\mathbf t}\,c_t^2+{\mathbf u}\,c_u^2\right)\notag\\
&&& \ \,-\Sigma_p\left(139\,\Sigma_p+27\right)\left[2\left({\mathbf s}\,\tilde{s}+{\mathbf t}\,\tilde{t}+{\mathbf u}\,\tilde{u}\right)+\left({\mathbf s}\,c_s+{\mathbf t}\,c_t+{\mathbf u}\,c_u\right)\right]\Big)\notag\\
&&+\left(\Sigma_p{{-}}1\right)_3\,& \Big(840\left(\tilde{s}^4+\tilde{t}^4+\tilde{u}^4\right)+\frac{191}{8}\left(c_s^4+c_t^4+c_u^4\right)+1680\left(\tilde{s}^3\,c_s+\tilde{t}^3\,c_t+\tilde{u}^3\,c_u\right)\notag\\
&&&\ \,+1140\left(\tilde{s}^2\,c_s^2+\tilde{t}^2\,c_t^2+\tilde{u}^2\,c_u^2\right)+300\left(c_s^3\,\tilde{s}+c_t^3\,\tilde{t}+c_u^3\,\tilde{u}\right)-960\,\Sigma_p\left(\tilde{s}^3+\tilde{t}^3+\tilde{u}^3\right)\notag\\
&&& \ \,-\frac{191}{2}\,\Sigma_p\left(c_s^3+c_t^3+c_u^3\right)-1440\,\Sigma_p\left(\tilde{s}^2\,c_s+\tilde{t}^2\,c_t+\tilde{u}^2\,c_u\right)\notag\\
&&&\ \,-671\,\Sigma_p\left(c_s^2\,\tilde{s}+c_t^2\,\tilde{t}+c_u^2\,\tilde{u}\right)+610\,\Sigma_p^2\left[\left(\tilde{s}^2+\tilde{t}^2+\tilde{u}^2\right)+\left(\tilde{s}\,c_s+\tilde{t}\,c_t+\tilde{u}\,c_u\right)\right]\notag\\
&&&\ \,+\frac{573}{4}\,\Sigma_p^2\left(c_s^2+c_t^2+c_u^2\right)-\frac{1471}{8}\,\Sigma_p^4-116\,\Sigma_p^2\Big)\Big]\ ,
\end{align}	
}and the Mellin amplitudes of the ambiguities are given in appendix~\ref{appendix:alpha7ambiguities}. Note that this exhibits a similar structure to \eqref{4dervmellinFA} and \eqref{6dervmellinFA}, since every line is multiplied by a Pochhammer depending on the power of $\{{\mathbf s},{\mathbf t},{\mathbf u}\}$ and the rest is at most quartic in the variables $\{{\mathbf s},{\mathbf t},{\mathbf u},\tilde{s},\tilde{t},\tilde{u},c_s,c_t,c_u,\Sigma_p\}$\, .

Collecting all possible contributions at this order, the complete Mellin amplitude for the $1/2$-BPS correlator at $\alpha'^7$ is given by eleven terms:
\begin{align}\label{ap7span}
\cM_{\alpha'^7}=\frac18\left(\frac{\alpha'}2\right)^7&\left(\tfrac{1}{2}\,\zeta_7 \mathcal{M}_{\alpha'^{7}}^{{\rm {main}}}+G_{1;0}\mathcal{M}_{\alpha'^{7}}^{{\rm {amb}_1}}+G_{2;0}\mathcal{M}_{\alpha'^{7}}^{{\rm {amb}_2}}+G_{3;0}\mathcal{M}_{\alpha'^{7}}^{{\rm {amb}_3}}+G_{4;0}\mathcal{M}_{\alpha'^{7}}^{{\rm {amb}_4}}+G_{5;0}\mathcal{M}_{\alpha'^{7}}^{{\rm {amb}_5}}\right.\notag\\
&\left.\ \ +D_1\mathcal{M}_{\alpha'^{6}}^{{\rm {main}}}+E_1\mathcal{M}_{\alpha'^{6}}^{{\rm {amb}}}+B_2\mathcal{M}_{\alpha'^{5}}^{{\rm {main}}}+C_2\mathcal{M}_{\alpha'^{5}}^{{\rm {amb}}}+A_4\mathcal{M}_{\alpha'^{3}}^\text{main}\right)\ .
\end{align}
The coefficients of the subleading terms remain unfixed at this order, to fix them we would need additional information. As an example, let us look at the lowest charge correlator with $p_i=2$ (as in the previous section we shift  $s\rightarrow\tfrac{s}{2}-2$\, , $t\rightarrow\tfrac{t}{2}-2$): 
\begin{align}\label{alpha72222}
\cM_{\alpha'^7}^{2222}=\frac18\left(\frac{\alpha'}2\right)^7\times60\,\Big(a_1\,\left(s^2 + t^2 + u^2\right)^2+a_2\,s\,t\,u \,+a_3\, \left(s^2 + t^2 + u^2\right)+ a_4\Big)\ ,
\end{align}
with $u=4-s-t$ and
\begin{align}
&a_1=1512\, \zeta_7 ,\quad a_2=336\,\left(D_1+48\left(G_{5;0}-2\,\zeta_7\right)\right) \ ,\notag\\
&a_3=42\, B_2+28\left(D_1-6\left(2\,E_1-18\,G_{1;0}-20\,G_{2;0}+40\,G_{3;0}-12\,G_{5;0}+23\, \zeta_7\right)\right)\ ,\notag\\
&a_4=A_4-32\left(3\,B_2+50\,D_1-12\left(2\,E_1-18\,G_{1;0}-20\,G_{2;0}+40\,G_{3;0}-204\,G_{5;0}+335\,\zeta_7\right)\right)\ .
\end{align}

\section{Conclusion} \label{conclusion}

In this paper, we postulate a simple effective field theory describing four-point tree-level string interactions in AdS$_5\times$S$^5$. Using a new formulation of Witten diagrams and the Mellin transform which treats AdS and S on equal footing, we show that this simple description reproduces previous results for all four-point correlators of $1/2$-BPS operators in $\mathcal{N}=4$ SYM up to order $\alpha'^5$, and propose a general algorithm for extending this to arbitrary high order, which makes new predictions at $\alpha'^6$ and $\alpha'^7$. The coefficients of the effective action can be determined by writing down an effective action which gives rise to the flat space VS amplitude and lifting it to AdS$_5\times$S$^5$, although there are curvature-dependent ambiguities which cannot be fixed in this way and need additional input from other methods such as localisation. After fixing all the coefficients in the effective action, the 10d Mellin amplitudes we derive from it can be thought of as the analogue of the VS amplitude in AdS$_5\times$S$^5$.

Note that here we have focused on the limit of tree-level string theory for which all orders in the $\alpha'$ effective action are known in flat space. However, the coefficients of the first three terms in the flat space effective action~\eqref{seff} (i.e. up to $\partial^6 \phi^4$) are actually known at the full non-perturbative level as functions of the string coupling~\cite{Green:1997as,Green:1998by,Green:1999pu,Green:2005ba,Green:2014yxa}.%
\footnote{We thank Congkao Wen for drawing our attention to this.} 
These results imply that  the coefficients in~\eqref{coeffexpansion} can be promoted to full functions of the complex Yang-Mills coupling $\tau=\theta/(2\pi)+4\pi i N \alpha'^2$ (where $\theta$ is the Yang-Mills theta-angle and we recall that $N \alpha'^2 = g_{YM}^{-2}$ if we set the AdS radius $R=1$). Specifically they are promoted as
\begin{align}
   \left(\tfrac{\alpha'}2\right)^3 A_0= \left(\tfrac{\alpha'}2\right)^32\zeta_3 \quad &\rightarrow \quad  \frac1{(2^4 \pi N)^{3/2}}\times E(\tfrac32,\tau,\bar \tau)\notag\\
   \left(\tfrac{\alpha'}2\right)^5 B_0= \left(\tfrac{\alpha'}2\right)^5 \zeta_5 \quad &\rightarrow \quad \frac1{(2^4 \pi N)^{5/2}}\times \tfrac12E(\tfrac52,\tau,\bar \tau)\notag\\
   \left(\tfrac{\alpha'}2\right)^6 D_0= \left(\tfrac{\alpha'}2\right)^6 2(\zeta_3)^2 \quad &\rightarrow \quad  \frac1{(2^4 \pi N)^{3}} \times 3\, {\mathcal E}(3,\tfrac32,\tfrac32,\tau,\bar \tau)
\label{tau}
\end{align}
where $A_0, B_0, D_0,$ are the leading coefficients in the first, second, and fourth lines of \eqref{coeffexpansion},
\begin{align}
    E(s,\tau,\bar \tau) = 2\zeta_{2s} (\Im(\tau))^s (1+\dots)
\end{align}
are non-holomorphic Eisenstein series and
\begin{align}
    {\mathcal E}(3,\tfrac32,\tfrac32,\tau,\bar \tau) = \tfrac23(\zeta_{3})^2 (\Im(\tau))^3 (1+\dots)
\end{align}
is a generalised Eisenstein series. 
In the above two equations the ellipses denote perturbative and non-perturbative terms which vanish when $\Im(\tau)\rightarrow \infty$. The precise definitions of the functions can be found for example in~\cite{Chester:2020vyz}.

Furthermore recently in~\cite{Chester:2019jas,Chester:2020vyz,Green:2020eyj} the
corresponding dual (but lowest charge only)  correlators were considered and completely fixed via localisation to all orders. This then fixes the remaining ambiguities at this order (assuming $B_1=C_1=0$ as we discuss around~\eqref{alpha62222} and the second bullet point below) in terms of the above functions as
\begin{align}
    C_0=-\frac{3}{2}B_0\ ,\qquad A_2=-30B_0\ , \qquad    E_0=\frac{D_0}{3}\ .
\end{align}
These relations follow from the earlier results in~\eqref{alpha5coeffs} and~\eqref{alpha6coeffs}, respectively.

In summary, the 10d effective action in~\eqref{Seff} appears to be a very useful way to describe IIB string theory in AdS$_5\times$S$^5$ and a powerful tool for computing four-point correlators in $\mathcal{N}=4$ SYM. There are a number of  interesting questions that remain:
\begin{itemize}

\item The strategy we have taken in this paper was to postulate a local 10d effective theory describing the string corrections to IIB supergravity in AdS$_5\times$S$^5$ and then use it to compute four-point correlators in $\mathcal{N}=4$ SYM. Just as on flat background~\cite{Howe:1983sra},  
IIB supergravity linearised on the AdS$\times$S superspace background
is again described by a chiral scalar superfield with a certain fourth order constraint~\cite{Heslop:2000np,heslopthesis}. 
 It presumably then makes sense to integrate a superpotential consisting of a holomorphic function of this scalar in chiral AdS$\times$S superspace. 
Ultimately however we do not derive the existence of this superpotential, but justify it a posteriori  by the fact that it reproduces previous results for correlators that were obtained using bootstrap methods~\cite{1809.10670,Binder:2019jwn,Drummond:2019odu,Drummond:2020dwr,Aprile:2020luw}. 
But this does then lead to the question of  the existence of an effective  chiral superpotential describing the full nonlinear theory. 
Such an object has been discussed before, notably  in~\cite{deHaro:2002vk,bh,Rajaraman:2005up}. An obstruction to its existence  was found in~\cite{bh,deHaro:2002vk}. However, despite this it was shown in~\cite{Rajaraman:2005up} that all terms  in the full non-linear effective action  consisting of the curvature and  five-form field strength
{\em are} correctly reproduced by such a superpotential. This latter fact perhaps explains the  existence of a superpotential in AdS$_5\times$S$^5$ (since the AdS$_5\times$S$^5$ superspace has only the five form field strength turned on). In any case it would be interesting to explore this point further.

\item As mentioned above, the effective action has ambiguities corresponding to curvature corrections which are invisible in the flat space limit. For low orders in $\alpha'$, we find that these ambiguities can be fixed by comparing to results from localisation. 
If it were possible to find a systematic way to fix all the ambiguities, the next question would be whether we can resum the $\alpha'$ expansion to obtain a compact form analogous to the flat space VS amplitude. If so, how does the analytic structure become modified in curved background?
Note here that, as observed below~\eqref{alpha62222},  the explicit results for these ambiguities obtained via localisation are completely consistent with all odd powers in the expansion of  $\alpha'/R^2$ vanishing.
Since the curvature  has opposite sign for AdS ($-20/R^2$)  and S ($+20/R^2$), it is perhaps quite natural to expect that only even powers of the curvature should contribute.
Moreover, we find that the nonzero coefficients which can be fixed by comparing to localisation results take a very simple form proportional to the leading coefficients. It would therefore be interesting to see if this simplicity or any other constraints can be derived from underlying symmetries of the superstring in AdS$_5 \times$S$^5$, such as the combination of bosonic and fermionic T-dualities discovered in~\cite{Berkovits:2008ic,Beisert:2008iq}.  
It should also be noted that the modular functions appearing in~\eqref{tau} are a consequence of the $SL(2,\mathbb{Z})$ symmetry of IIB string theory, which can be understood from compactifying M-theory on a torus and identifying the IIB coupling $\tau$ with the complex structure of the torus~\cite{Aspinwall:1995fw,Schwarz:1995jq}.

\item It would also be interesting to extend this approach to higher-point correlators. The four-point AdS$\times$S contact diagrams have a direct generalisation to $n$ points.
An important feature of $1/2$-BPS four-point correlators in $\mathcal{N}=4$ SYM that allowed us to write down a simple effective action was the ability to factor out a polynomial which encodes all the supersymmetry.
This is analogous to factoring out a supersymmetric delta function $\delta^{16}(Q)$ from a maximally supersymmetric four-point superamplitude in flat space. A similar property holds for $n$-point maximally nilpotent correlators -- those with fermionic degree $n-4$~\cite{hep-th/9905085,hep-th/9910011,1108.3557}. These have  recently been studied at strong coupling in~\cite{Green:2020eyj} and one might  expect them to be computable from a 10d scalar effective action just as for the four point ones. 

\item Another important direction would be to extend this approach to other backgrounds, in particular other backgrounds of the form AdS$_q\times$S$^q$. For $q=3,5$, it was recently shown that supergravity correlators enjoy conformal symmetry which can be used to lift the lowest charge $1/2$-BPS four-point correlator to all higher charge correlators~\cite{Caron-Huot:2018kta,Rastelli:2019gtj,Giusto:2020neo}. 
It would be interesting to investigate the relation of this higher dimensional conformal symmetry with the explicit higher dimensional integrals (AdS$\times$S Witten diagrams) we write down here.
While the approach in the above references is restricted to supergravity, ours naturally describes string corrections. However 
note that the results of~\cite{Drummond:2019odu,Drummond:2020dwr,Aprile:2020luw} strongly hint  that much of the higher dimensional conformal structure survives for the string corrections.
It would therefore be interesting to investigate how these two approaches are related.

\item 
It would also be conceptually very satisfying to {\em derive} the effective action directly from CFT without assuming local spacetime description in the bulk. Indeed, deriving the emergence of spacetime from a non-gravitational theory in the boundary is one of the main reasons for the huge interest in AdS/CFT. 
This is also essentially the goal of the bootstrap approach~\cite{Drummond:2019odu,Drummond:2020dwr,Aprile:2020luw}.
A systematic approach to such a derivation was achieved in the context of a toy model consisting of a scalar field in AdS in~\cite{Heemskerk:2009pn} using crossing and conformal symmetry of boundary CFT correlators. This calculation was adapted to stress tensor correlators in $\mathcal{N}=4$ SYM in~\cite{Alday:2014tsa}. The fact that IIB string theory in AdS$_5\times$S$^5$ can be reduced to a simple 10d effective field theory therefore suggests the exciting prospect that this program might be realised for a full-blown theory of quantum gravity. This would presumably require better understanding of the relation to the higher dimensional conformal symmetry discussed above.

\item Finally there is the important question of performing loop computations of amplitudes directly on the gravitational side of the duality (rather than bootstrapping loop results from the CFT side as has been achieved in  recent years~\cite{1706.02822,1711.02031,1711.03903,1912.01047,1912.02663,1809.10670,1812.11783,1912.07632,2008.01109}). One of the obstacles to performing loop computations directly on the gravity side is the sheer technical complexity of summing over all the supergravity fields and one could hope there is a way of leveraging the tree-level simplicity we have uncovered here to help at loop level.

\end{itemize}

We hope to report further on these directions in the future.

\begin{center}
\textbf{Acknowledgements}
\end{center}
We would like to thank Francesco Aprile, James Drummond, Hynek Paul, Michele Santagata and Congkao Wen for insightful discussions.
TA is supported by a Durham Doctoral Studentship, AL by the Royal Society as a Royal Society University Research Fellowship holder, and PH by an STFC Consolidated Grant ST/P000371/1. We are also part of the
European Union's Horizon 2020 research and innovation programme 
under the Marie Sk\l{}odowska-Curie grant agreement No. 764850 ``SAGEX''.

\appendix

\section{The polynomial $I(X_i,Y_i)$}
\label{intriligator}

The polynomial $I(X_i,Y_i)$ which factors out of all $1/2$-BPS four-point functions~\eqref{4intdef} is:
\begin{align} 
		I(X_i,Y_i)&= (x-y)(\bar x -y)(x-\bar y)(\bar x - \bar y)(X_{1}.X_{3})^2 (X_{2}.X_{4})^2(Y_{1}.Y_{3})^2 (Y_{2}.Y_{4})^2\notag\\
		x \bar x &= \frac{X_{1}.X_{2}X_{3}.X_{4}}{X_{1}.X_{3}X_{2}.X_{4}} \qquad (1-x) (1-\bar x) = \frac{X_{1}.X_{4}X_{2}.X_{3}}{X_{1}.X_{3}X_{2}.X_{4}}\notag \\
		y \bar y &= \frac{Y_{1}.Y_{2}Y_{3}.Y_{4}}{Y_{1}.Y_{3}Y_{2}.Y_{4}} \qquad (1-y) (1-\bar y) = \frac{Y_{1}.Y_{4}Y_{2}.Y_{3}}{Y_{1}.Y_{3}Y_{2}.Y_{4}}\ .
	\end{align}
	$I$ is crossing symmetric, under simultaneously permuting $X_i,Y_i$ with $X_j,Y_j$. It is also a polynomial and written out fully in terms of the $SO(2,4)$ and $SO(6)$ invariants $X_i.X_j$ and $Y_i.Y_j$ is given as
	\begin{align}	&I(X_i,Y_i)=\notag\\
		&	\left(X_1.X_4\right){}^2 \left(X_2.X_3\right){}^2 Y_1.Y_2 Y_1.Y_3 Y_2.Y_4 Y_3.Y_4+X_1.X_2 X_1.X_4 X_3.X_4 X_2.X_3 \left(Y_1.Y_3\right){}^2
		\left(Y_2.Y_4\right){}^2\notag\\
		&+X_1.X_3 X_1.X_4 X_2.X_4 X_2.X_3 \left(Y_1.Y_2\right){}^2 \left(Y_3.Y_4\right){}^2-X_1.X_2 X_1.X_4 X_3.X_4 X_2.X_3
		Y_1.Y_3 Y_1.Y_4 Y_2.Y_3 Y_2.Y_4\notag\\
		&-X_1.X_3 X_1.X_4 X_2.X_4 X_2.X_3 Y_1.Y_2 Y_1.Y_4 Y_2.Y_3 Y_3.Y_4-X_1.X_3 X_1.X_4 X_2.X_4 X_2.X_3 Y_1.Y_2
		Y_1.Y_3 Y_2.Y_4 Y_3.Y_4\notag\\
		&-X_1.X_2 X_1.X_4 X_3.X_4 X_2.X_3 Y_1.Y_2 Y_1.Y_3 Y_2.Y_4 Y_3.Y_4+X_1.X_2 X_1.X_3 X_2.X_4 X_3.X_4
		\left(Y_1.Y_4\right){}^2 \left(Y_2.Y_3\right){}^2\notag\\
		&+\left(X_1.X_2\right){}^2 \left(X_3.X_4\right){}^2 Y_1.Y_3 Y_1.Y_4 Y_2.Y_3
		Y_2.Y_4-X_1.X_2 X_1.X_3 X_2.X_4 X_3.X_4 Y_1.Y_3 Y_1.Y_4 Y_2.Y_3 Y_2.Y_4\notag\\
		&+\left(X_1.X_3\right){}^2 \left(X_2.X_4\right){}^2 Y_1.Y_2 Y_1.Y_4
		Y_2.Y_3 Y_3.Y_4-X_1.X_2 X_1.X_3 X_2.X_4 X_3.X_4 Y_1.Y_2 Y_1.Y_4 Y_2.Y_3 Y_3.Y_4
		\ .
	\end{align}

\section{$\alpha'^7$ ambiguities}\label{appendix:alpha7ambiguities}

The ambiguities at order $\alpha'^7$ were introduced in~\eqref{alpha7ambs} and we spell out their Witten diagram expressions and the corresponding Mellin amplitudes in the following.

The first ambiguity at $\alpha'^7$ contributes to the effective action with
\begin{align}\label{8derivamb1}
S_{\alpha'^7}^{\text{amb}_1} &=  \frac6{4!}\int_{\text{AdS}\times \text{S}} d^5\hat X d^5 \hat Y    \left(\nabla^\mu\nabla^\nu\nabla_\mu\nabla^\rho\nabla^\sigma\nabla_\rho\phi\right)\left(\nabla_\nu\nabla_\sigma\phi\right)\phi^2,
\end{align}
it corresponds to a four-derivative interaction and its contribution to the $1/2$-BPS correlator is given as
\begin{align}\label{ap8amb1}
\langle \cO\cO\cO\cO\rangle|_{\alpha'^7;\text{amb}_1}=\frac1{4!}\frac {(\mathcal{C}_4)^4}{(-2)^{16}} \int_{\text{AdS}\times\text{S}}\frac{d^{5}\hat{X}d^{5}\hat{Y}}{\prod_i \left(P_{i}+Q_{i}\right)^{4}}\sum_{i<j}\frac{K^{\text{amb}_1}_{ij}}{\left(P_{i}+Q_{i}\right)^2\left(P_{j}+Q_{j}\right)^2}\times 4^3\times 5\ ,
 \end{align} 
where
\begin{align}\label{Kamb1}
K_{ij}^{\text{amb}_1}=&\,45 \left[\left(X_i.X_j+P_i P_j\right)^2+\left(Y_i.Y_j+Q_i Q_j\right)^2\right]-9\left(P_i P_j +Q_i Q_j\right)^2\notag\\
&-180\, Q_i Q_j\, Y_i.Y_j+26\,P_i P_j\, Q_i Q_j\ .
\end{align}
We write the corresponding Mellin amplitude as
\begin{align}
\cM_{\alpha'^7}^{\text{amb}_1}=\hat{\cM}_{\alpha'^7}^{\text{amb}_1}+204 \cM_{\alpha'^5}^\text{amb}+12288 \cM_{\alpha'^3}^\text{main}\ ,
\end{align}
where
{\small
\begin{align}
\hat{\mathcal{M}}_{\alpha'^{7}}^{{\rm {amb}}_1}=&& 288\,\Big[\left(\Sigma_p{{-}}1\right)_5\,&\left({\mathbf s}^2+{\mathbf t}^2+{\mathbf u}^2\right)\notag\\
&& +\left(\Sigma_p{{-}}1\right)_4\,&\left(\frac{1}{2}\left({\mathbf s}\,c_s^2+{\mathbf t}\,c_t^2+{\mathbf u}\,c_u^2\right)-\left(\Sigma_p+3\right)\left[2\left({\mathbf s}\,\tilde{s}+{\mathbf t}\,\tilde{t}+{\mathbf u}\,\tilde{u}\right)+\left({\mathbf s}\,c_s+{\mathbf t}\,c_t+{\mathbf u}\,c_u\right)\right]\right)\notag\\
&& +\left(\Sigma_p{{-}}1\right)_3\,&\Big(\frac{1}{12}\left(c_s^4+c_t^4+c_u^4\right)-\Sigma_p\left(c_s^2\,\tilde{s}+c_t^2\,\tilde{t}+c_u^2\,\tilde{u}^2\right)+\frac{1}{18}\left(c_s^2\,c_t^2+c_s^2\,c_u^2+c_t^2\,c_u^2\right)\notag\\
&&&\ \,-\frac{1}{2}\Sigma_p\left(c_s^3+c_t^3+c_u^3\right)+\frac{29}{36}\Sigma_p^2\left(c_s^2+c_t^2+c_u^2\right)-\frac{5}{6}\,c_s\,c_t\,c_u\,\Sigma_p\notag\\
&&&\ \,+2\left(\Sigma_p^2+6\right)\left[\left(\tilde{s}^2+\tilde{t}^2+\tilde{u}^2\right)+\left(\tilde{s}\,c_s+\tilde{t}\,c_t+\tilde{u}\,c_u\right)\right]-\frac{1}{6}\Sigma_p^2\left(\Sigma_p^2+72\right)\Big)\Big]\ .
\end{align}
}
The next three ambiguities also correspond to four-derivative terms. The contribution to the effective action from the second ambiguity is
\begin{align}\label{8derivamb2}
S_{\alpha'^7}^{\text{amb}_2} =  \frac6{4!}\int_{\text{AdS}\times \text{S}} d^5\hat X d^5 \hat Y    \left(\nabla^2\nabla^\mu\nabla^\nu\nabla^\rho\nabla_\nu\phi\right)\left(\nabla_\mu\nabla_\rho\phi\right)\phi^2,
\end{align}
which corresponds to the correlator
\begin{align}\label{ap8amb2}
\langle \cO\cO\cO\cO\rangle|_{\alpha'^7;\text{amb}_2}=\frac1{4!}\frac {(\mathcal{C}_4)^4}{(-2)^{16}} \int_{\text{AdS}\times\text{S}}\frac{d^{5}\hat{X}d^{5}\hat{Y}}{\prod_i \left(P_{i}+Q_{i}\right)^{4}}\sum_{i<j}\frac{K^{\text{amb}_2}_{ij}}{\left(P_{i}+Q_{i}\right)^2\left(P_{j}+Q_{j}\right)^2}\times 4^3\times 5^2\times 2\ ,
 \end{align} 
where
\begin{align}\label{Kamb2}
K_{ij}^{\text{amb}_2}=&\,5 \left[\left(X_i.X_j+P_i P_j\right)^2+\left(Y_i.Y_j+Q_i Q_j\right)^2\right]-\left(P_i P_j +Q_i Q_j\right)^2\notag\\
&-20\, Q_i Q_j\, Y_i.Y_j+2\,P_i P_j\, Q_i Q_j\ .
\end{align}
The Mellin amplitude is
\begin{align}
\cM_{\alpha'^7}^{\text{amb}_2}=\hat{\cM}_{\alpha'^7}^{\text{amb}_2}+248 \cM_{\alpha'^5}^\text{amb}+14336 \cM_{\alpha'^3}^\text{main}\ ,
\end{align}
where
{\small
\begin{align}
\hat{\mathcal{M}}_{\alpha'^{7}}^{{\rm {amb}}_2}=&& 320\Big[ \left(\Sigma_p{{-}}1\right)_5\,&\left({\mathbf s}^2+{\mathbf t}^2+{\mathbf u}^2\right)\notag\\
&& +\left(\Sigma_p{{-}}1\right)_4\,&\left(\frac{1}{2}\left({\mathbf s}\,c_s^2+{\mathbf t}\,c_t^2+{\mathbf u}\,c_u^2\right)-\left(\Sigma_p+3\right)\left[2\left({\mathbf s}\,\tilde{s}+{\mathbf t}\,\tilde{t}+{\mathbf u}\,\tilde{u}\right)+\left({\mathbf s}\,c_s+{\mathbf t}\,c_t+{\mathbf u}\,c_u\right)\right]\right)\notag\\
&& +\left(\Sigma_p{{-}}1\right)_3\,&\Big(\frac{3}{40}\left(c_s^4+c_t^4+c_u^4\right)-\Sigma_p\left(c_s^2\,\tilde{s}+c_t^2\,\tilde{t}+c_u^2\,\tilde{u}^2\right)+\frac{1}{20}\left(c_s^2\,c_t^2+c_s^2\,c_u^2+c_t^2\,c_u^2\right)\notag\\
&&&\ \,-\frac{1}{2}\Sigma_p\left(c_s^3+c_t^3+c_u^3\right)+\frac{4}{5}\Sigma_p^2\left(c_s^2+c_t^2+c_u^2\right)-\frac{9}{10}\,c_s\,c_t\,c_u\,\Sigma_p\notag\\
&&&\ \,+2\left(\Sigma_p^2+6\right)\left[\left(\tilde{s}^2+\tilde{t}^2+\tilde{u}^2\right)+\left(\tilde{s}\,c_s+\tilde{t}\,c_t+\tilde{u}\,c_u\right)\right]-\frac{1}{40}\Sigma_p^2\left(7\,\Sigma_p^2+480\right)\Big)\Big]\ .
\end{align}
}

The third ambiguity contributes to the effective action with
\begin{align}\label{8derivamb3}
S_{\alpha'^7}^{\text{amb}_3} &=  \int_{\text{AdS}\times \text{S}} d^5\hat X d^5 \hat Y    \left(\nabla^2\nabla^\mu\nabla^\nu\nabla^\rho\nabla_\nu\phi\right)\left(\nabla_\rho\phi\right)\left(\nabla_\mu\phi\right)\phi,
\end{align}
and the prediction for its contribution to the $1/2$-BPS correlator is given by
\begin{align}\label{ap8amb3}
&\langle \cO\cO\cO\cO\rangle|_{\alpha'^7;\text{amb}_3}=\notag\\
&\frac1{4!}\frac {(\mathcal{C}_4)^4}{(-2)^{16}} \int_{\text{AdS}\times\text{S}}\frac{d^{5}\hat{X}d^{5}\hat{Y}}{\prod_i \left(P_{i}+Q_{i}\right)^{4}}\left[\frac{K^{\text{amb}_3}_{123}}{\left(P_{1}+Q_{1}\right)^2\left(P_{2}+Q_{2}\right)\left(P_{3}+Q_{3}\right)}+\text{perms}\right]\times 4^4\times 5\times 2\ ,
 \end{align} 
where we sum over all permutations and
\begin{align}\label{Kamb3}
K_{ijk}^{\text{amb}_3}=&\,P_i^2 \left(4\,P_j P_k-X_j.X_k\right)+Q_i^2 \left(4\, Q_j Q_k+Y_j.Y_k\right)+5\,P_i\left( P_k\, X_i.X_j+ P_j\, X_i.X_k\right)\notag\\
&-5\,Q_i \left( Q_k\, Y_i.Y_j+ Q_j\, Y_i.Y_k\right)+5 \left(X_i.X_j\, X_i.X_k+Y_i.Y_j\, Y_i.Y_k\right)\ .
\end{align}
The corresponding Mellin amplitude is
\begin{align}
\cM_{\alpha'^7}^{\text{amb}_3}=\hat{\cM}_{\alpha'^7}^{\text{amb}_3}-704 \cM_{\alpha'^5}^\text{amb}-32768 \cM_{\alpha'^3}^\text{main}\ ,
\end{align}
where
{\small
\begin{align}
\hat{\mathcal{M}}_{\alpha'^{7}}^{{\rm {amb}}_3}=&&-640\Big[ \left(\Sigma_p{{-}}1\right)_5\,&\left({\mathbf s}^2+{\mathbf t}^2+{\mathbf u}^2\right)\notag\\
&& +\left(\Sigma_p{{-}}1\right)_4\,&\left(\frac{1}{2}\left({\mathbf s}\,c_s^2+{\mathbf t}\,c_t^2+{\mathbf u}\,c_u^2\right)-\left(\Sigma_p+3\right)\left[2\left({\mathbf s}\,\tilde{s}+{\mathbf t}\,\tilde{t}+{\mathbf u}\,\tilde{u}\right)+\left({\mathbf s}\,c_s+{\mathbf t}\,c_t+{\mathbf u}\,c_u\right)\right]\right)\notag\\
&& +\left(\Sigma_p{{-}}1\right)_3\,&\Big(\frac{1}{20}\left(c_s^4+c_t^4+c_u^4\right)-\Sigma_p\left(c_s^2\,\tilde{s}+c_t^2\,\tilde{t}+c_u^2\,\tilde{u}^2\right)-\frac{1}{10}\left(c_s^2\,c_t^2+c_s^2\,c_u^2+c_t^2\,c_u^2\right)\notag\\
&&&\ \,-\frac{1}{2}\Sigma_p\left(c_s^3+c_t^3+c_u^3\right)+\frac{13}{20}\Sigma_p^2\left(c_s^2+c_t^2+c_u^2\right)-\frac{3}{10}\,c_s\,c_t\,c_u\,\Sigma_p\notag\\
&&&\ \,+2\left(\Sigma_p^2+6\right)\left[\left(\tilde{s}^2+\tilde{t}^2+\tilde{u}^2\right)+\left(\tilde{s}\,c_s+\tilde{t}\,c_t+\tilde{u}\,c_u\right)\right]-\frac{1}{5}\Sigma_p^2\left(\Sigma_p^2+60\right)\Big)\Big]\ .
\end{align}
}

The next ambiguity contributes to the effective action as
\begin{align}\label{8derivamb4}
S_{\alpha'^7}^{\text{amb}_4} &=  \int_{\text{AdS}\times \text{S}} d^5\hat X d^5 \hat Y    \left(\nabla^\mu\nabla^\nu\nabla^\rho\nabla_\nu\nabla_\rho\phi\right)\left(\nabla^\sigma\nabla_\mu\phi\right)\left(\nabla_\sigma\phi\right)\phi,
\end{align}
and the corresponding correlator is given by
\begin{align}\label{ap8amb4}
&\langle \cO\cO\cO\cO\rangle|_{\alpha'^7;\text{amb}_4}=\notag\\
&\frac1{4!}\frac {(\mathcal{C}_4)^4}{(-2)^{16}} \int_{\text{AdS}\times\text{S}}\frac{d^{5}\hat{X}d^{5}\hat{Y}}{\prod_{i}\left(P_{i}+Q_{i}\right)^{4}}\left[\frac{K^{\text{amb}_4}_{123}}{\left(P_{1}+Q_{1}\right)^3\left(P_{2}+Q_{2}\right)^2\left(P_{3}+Q_{3}\right)}+\text{perms}\right]\times 4^4\times 5\times 2\ ,
 \end{align} 
where
\begin{align}\label{Kamb4}
K_{ijk}^{\text{amb}_4}=&\,P_i^2 Q_i \left[5 \left(5 P_j+Q_j\right) \left(X_j.X_k+Y_j.Y_k\right)-4 Q_j Q_k \left(6 P_j+Q_j\right)+20 P_j^2 P_k\right]\notag\\
&-P_i Q_i^2 \left[5 \left(P_j+5 Q_j\right) \left(X_j.X_k+Y_j.Y_k\right)+4 P_j P_k \left(P_j+6 Q_j\right)-20 Q_j^2 Q_k\right]\notag\\
&-P_i^2 \left[Q_j \left(P_j+Q_j\right) Y_i.Y_k+5 Y_i.Y_j \left(X_j.X_k+Y_j.Y_k+P_j P_k-Q_j Q_k\right)\right]\notag\\
&+Q_i^2 \left[P_j \left(P_j+Q_j\right) X_i.X_k-5 X_i.X_j\left(X_j.X_k+Y_j.Y_k+P_j P_k-Q_j Q_k\right)\right]\notag\\
&+P_i Q_i \left[25 \left(P_j P_k-Q_j Q_k\right) \left(X_i.X_j+Y_i.Y_j\right)+\left(P_j+Q_j\right) \left(5 Q_j\, Y_i.Y_k-5 P_j\,X_i.X_k\right)\right.\notag\\
&\qquad\quad\ +25 \left.\left(X_i.X_j+Y_i.Y_j\right) \left(X_j.X_k+Y_j.Y_k\right)\right]\ .
\end{align}
The contribution of this ambiguity to the Mellin amplitude is
\begin{align}
\cM_{\alpha'^7}^{\text{amb}_4}=\hat{\cM}_{\alpha'^7}^{\text{amb}_4}-128 \cM_{\alpha'^5}^\text{main}\ ,
\end{align}
where 
{\small
\begin{align}
\hat{\mathcal{M}}_{\alpha'^{7}}^{{\rm {amb}}_4}=&& 32\,\Big[ \left(\Sigma_p{{-}}1\right)_5\,&\Big(\Sigma_p^2\left({\mathbf s}^2+{\mathbf t}^2+{\mathbf u}^2\right)+\left({\mathbf s}^2\,c_s^2+{\mathbf t}^2\,c_t^2+{\mathbf u}^2\,c_u^2\right)\notag\\
&&&\ \,+\left[{\mathbf s}^2\left(c_t^2+c_u^2\right)+{\mathbf t}^2\left(c_s^2+c_u^2\right)+{\mathbf u}^2\left(c_s^2+c_t^2\right)\right]\Big)\notag\\
&& +\left(\Sigma_p{{-}}1\right)_4\,&\Big(-5\left({\mathbf s}\,c_s^3+{\mathbf t}\,c_t^3+{\mathbf u}\,c_u^3\right)-10\left({\mathbf s}\,c_s^2\,\tilde{s}+{\mathbf t}\,c_t^2\,\tilde{t}+{\mathbf u}\,c_u^2\,\tilde{u}\right)-10\,\Sigma_p^2\left({\mathbf s}\,\tilde{s}+{\mathbf t}\,\tilde{t}+{\mathbf u}\,\tilde{u}\right)\notag\\
&&&\ \,-5\,\Sigma_p^2\left({\mathbf s}\,c_s+{\mathbf t}\,c_t+{\mathbf u}\, c_u\right)-10\left[{\mathbf s}\,\tilde{s}\left(c_t^2+c_u^2\right)+{\mathbf t}\,\tilde{t}\left(c_s^2+c_u^2\right)+{\mathbf u}\,\tilde{u}\left(c_s^2+c_t^2\right)\right]\notag\\
&&&\ \,-5\left[{\mathbf s}\,c_s\left(c_t^2+c_u^2\right)+{\mathbf t}\,c_t\left(c_s^2+c_u^2\right)+{\mathbf u}\,c_u\left(c_s^2+c_t^2\right)\right]\Big)\notag\\
&& +\left(\Sigma_p{{-}}1\right)_3\,&\Big(4\left(c_s^4+c_t^4+c_u^4\right)+20\left(\tilde{s}^2\,c_s^2+\tilde{t}^2\,c_t^2+\tilde{u}^2\,c_u^2\right)+8\left(c_s^2\,c_t^2+c_s^2\,c_u^2+c_t^2\,c_u^2\right)\notag\\
&&&\ \,+20\left(c_s^3\,\tilde{s}+c_t^3\,\tilde{t}+c_u^3\,\tilde{u}\right)+20\,\Sigma_p^2\left(\tilde{s}^2+\tilde{t}^2+\tilde{u}^2\right)-8\,\Sigma_p^2\left(c_s^2+c_t^2+c_u^2\right)\notag\\
&&&\ \,+20\,\Sigma_p^2\left(\tilde{s}\,c_s+\tilde{t}\,c_t+\tilde{u}\,c_u\right)+20\left[\tilde{s}^2\left(c_t^2+c_u^2\right)+\tilde{t}^2\left(c_s^2+c_u^2\right)+\tilde{u}^2\left(c_s^2+c_t^2\right)\right]\notag\\
&&&\ \,+20\left[\tilde{s}\,c_s\left(c_t^2+c_u^2\right)+\tilde{t}\,c_t\left(c_s^2+c_u^2\right)+\tilde{u}\,c_u\left(c_s^2+c_t^2\right)\right]-12\,\Sigma_p^4\Big)\Big]\ .
\end{align}
}

Finally, the fifth ambiguity contributes to the effective action with
\begin{align}\label{8derivamb5}
S_{\alpha'^7}^{\text{amb}_5} &= \int_{\text{AdS}\times \text{S}} d^5\hat X d^5 \hat Y    \left(\nabla^\mu\nabla^\nu\nabla^\rho\nabla^\sigma\nabla_\rho\phi\right)\left(\nabla_\mu\nabla_\sigma\phi\right)\left(\nabla_\nu\phi\right)\phi\ ,
\end{align}
which corresponds to a six-derivative interaction and its contribution to the $1/2$-BPS correlator is
\begin{align}\label{ap8amb5}
&\langle \cO\cO\cO\cO\rangle|_{\alpha'^7;\text{amb}_5}=\notag\\
&\frac1{4!}\frac {(\mathcal{C}_4)^4}{(-2)^{16}} \int_{\text{AdS}\times\text{S}}\frac{d^{5}\hat{X}d^{5}\hat{Y}}{\prod_{i}\left(P_{i}+Q_{i}\right)^{4}}\left[\frac{K^{\text{amb}_5}_{123}}{\left(P_{1}+Q_{1}\right)^3\left(P_{2}+Q_{2}\right)^2\left(P_{3}+Q_{3}\right)}+\text{perms}\right]\times (-4^4)\ ,
 \end{align} 
where
\begin{align}\label{Kamb5}
K_{ijk}^{\text{amb}_5}=&\,P_i^3 \left[P_j \left(4 \left( Q_j+21 P_j \right)P_k  - 45 X_j.X_k\right)\right]+Q_i^3 \left[Q_j \left(4 \left(P_j + 21 Q_j\right) Q_k + 45 Y_j.Y_k\right)\right]\notag\\
& + P_i^2 Q_i \left[8\left(Q_j-4 P_j\right)P_j P_k + 
    4 \left(Q_j-5 P_j\right)\left(Q_j+6 P_j\right) Q_k - 
    40 P_j X_j.X_k - 5 Q_j Y_j.Y_k\right]\notag\\
&+ P_i Q_i^2 \left[ 8 \left(P_j - 4 Q_j\right) Q_j Q_k +4  \left(P_j - 5 Q_j\right) \left(P_j + 6 Q_j\right)P_k  + 
    40 Q_j Y_j.Y_k+ 5 P_j X_j.X_k\right]\notag\\
& + P_i^2 \left[P_j \left(129 P_j + 4 Q_j\right) X_i.X_k + 
    15 X_i.X_j \left(17 P_j P_k - 3 X_j.X_k\right) + 
    120 P_j^2 Y_i.Y_k \right.\notag\\
&\qquad\ \ +\left. Q_j \left(-5 Q_k Y_i.Y_j + \left(-4 P_j + Q_j\right) Y_i.Y_k\right) +5 Y_i.Y_j Y_j.Y_k\right]\notag\\
& + Q_i^2 \left[-Q_j \left(129 Q_j + 4 P_j\right) Y_i.Y_k - 
    15 Y_i.Y_j \left(17 Q_j Q_k + 3 Y_j.Y_k\right) - 
    120 Q_j^2 X_i.X_k \right.\notag\\
&\qquad\ \ -\left. P_j \left(-5 P_k X_i.X_j + \left(-4 Q_j + P_j\right) X_i.X_k\right) +5 X_i.X_j X_j.X_k\right]\notag\\
&+P_i Q_i \left[8\left(P_j + Q_j\right) \left( P_j X_i.X_k -  Q_j Y_i.Y_k\right) \right.\notag\\
&\left.\qquad \quad +  20 \left(-X_i.X_j \left(P_j \left(2 P_k + 15 Q_k\right) + 2 X_j.X_k\right)+ Y_i.Y_j \left(Q_j \left( 2 Q_k+15 P_k\right) - 2 Y_j.Y_k\right)\right)\right]\notag\\
&+ P_i \left[150 P_k \left(\left(X_i.X_j\right)^2 - \left(Y_i.Y_j\right)^2\right) + 
    300 P_j X_i.X_j \left(X_i.X_k + Y_i.Y_k\right)\right]\notag\\
&+Q_i \left[-150 Q_k\left(\left(X_i.X_j\right)^2 -\left(Y_i.Y_j\right)^2\right)+300 Q_j Y_i.Y_j\left(X_i.X_k+Y_i.Y_k\right)\right]\notag\\
&+ 150 \left(\left(X_i.X_j\right)^2 - \left(Y_i.Y_j\right)^2\right) \left(X_i.X_k + Y_i.Y_k\right)\ .
\end{align}
The corresponding Mellin amplitude is
\begin{align}
\cM_{\alpha'^7}^{\text{amb}_5}=&\hat{\cM}_{\alpha'^7}^{\text{amb}_5}-\frac{11}{2}\cM_{\alpha'^7}^{\text{amb}_1}+5\cM_{\alpha'^7}^{\text{amb}_2}+\frac{1}{8}\cM_{\alpha'^7}^{\text{amb}_3}-\cM_{\alpha'^7}^{\text{amb}_4}+64 \cM_{\alpha'^5}^\text{main}\notag\\
&+66\cM_{\alpha'^5}^{\text{amb}}+4096\cM_{\alpha'^3}^\text{main}\ ,
\end{align}
where
{\small
\begin{align}
\hat{\mathcal{M}}_{\alpha'^{7}}^{{\rm {amb}}_5}=&& 128 \,\Big[ \left(\Sigma_p{{-}}1\right)_6\,&\left({\mathbf s}^3+{\mathbf t}^3+{\mathbf u}^3\right)\notag\\
&&+\left(\Sigma_p{{-}}1\right)_5\,&\Big(\frac{1}{2}\left({\mathbf s}^2\,c_s^2+{\mathbf t}^2\,c_t^2+{\mathbf u}^2\,c_u^2\right)+\frac{1}{2}\,\Sigma_p\left(\Sigma_p+8\right)\left({\mathbf s}^2+{\mathbf t}^2+{\mathbf u}^2\right)\notag\\
&&&\ \,-\left(\Sigma_p+7\right)\left[2\left({\mathbf s}^2\,\tilde{s}+{\mathbf t}^2\,\tilde{t}+{\mathbf u}^2\,\tilde{u}\right)+\left({\mathbf s}^2\,c_s+{\mathbf t}^2\,c_t+{\mathbf u}^2\,c_u\right)\right]\Big)\notag\\
&& +\left(\Sigma_p{{-}}1\right)_4\,&\Big(-\frac{5}{2}\left({\mathbf s}\,c_s^3+{\mathbf t}\,c_t^3+{\mathbf u}\,c_u^3\right)-5\left({\mathbf s}\,c_s^2\,\tilde{s}+{\mathbf t}\,c_t^2\,\tilde{t}+{\mathbf u}\,c_u^2\,\tilde{u}\right)\notag\\
&&&\ \,+6\left(4\,\Sigma_p+7\right)\left[\left({\mathbf s}\,\tilde{s}^2+{\mathbf t}\,\tilde{t}^2+{\mathbf u}\,\tilde{u}^2\right)+\left({\mathbf s}\,\tilde{s}\,c_s+{\mathbf t}\,\tilde{t}\,c_t+{\mathbf u}\,\tilde{u}\,c_u\right)\right]\notag\\
&&&\ \,-\Sigma_p\left(13\,\Sigma_p+24\right)\left[\left({\mathbf s}\,\tilde{s}+{\mathbf t}\,\tilde{t}+{\mathbf u}\,\tilde{u}\right)+\frac{1}{2}\left({\mathbf s}\,c_s+{\mathbf t}\,c_t+{\mathbf u}\, c_u\right)\right]\notag\\
&&&\ \,+\frac{1}{2}\left(14\,\Sigma_p+19\right)\left({\mathbf s}\,c_s^2+{\mathbf t}\,c_t^2+{\mathbf u}\,c_u^2\right)\Big)\notag\\
&& +\left(\Sigma_p{{-}}1\right)_3\,&\Big(\frac{17}{8}\left(c_s^4+c_t^4+c_u^4\right)-60\,\Sigma_p\left(\tilde{s}^3+\tilde{t}^3+\tilde{u}^3\right)-\frac{17}{2}\,\Sigma_p\left(c_s^3+c_t^3+c_u^3\right)\notag\\
&&&\ \,+10\left(c_s^3\,\tilde{s}+c_t^3\,\tilde{t}+c_u^3\,\tilde{u}\right)+10\left(\tilde{s}^2\,c_s^2+\tilde{t}^2\,c_t^2+\tilde{u}^2\,c_u^2\right)-90\,\Sigma_p\left(\tilde{s}^2\,c_s+\tilde{t}^2\,c_t+\tilde{u}^2\,c_u\right)\notag\\
&&&\ \,-47\,\Sigma_p\left(c_s^2\,\tilde{s}+c_t^2\,\tilde{t}+c_u^2\,\tilde{u}\right)+\frac{51}{4}\,\Sigma_p^2\left(c_s^2+c_t^2+c_u^2\right)+50\,\Sigma_p^2\left(\tilde{s}^2+\tilde{t}^2+\tilde{u}^2\right)\notag\\
&&&\ \,+50\,\Sigma_p^2\left(\tilde{s}\,c_s+\tilde{t}\,c_t+\tilde{u}\,c_u\right)-\frac{1}{8}\,\Sigma_p^2\left(81\,\Sigma_p^2+352\right)\Big)\Big]\ .
\end{align}	
}

\end{document}